\documentclass[12pt]{report}
\usepackage{harvard}
\citationstyle{dcu} \textwidth=14cm \textheight=23cm
\usepackage{amsfonts}
\usepackage{epsfig}
\usepackage{lscape}

\begin{document}
\begin{center}

\LARGE \textbf{Angular Momentum Imparted \\To Test Particles by\\
Gravitational Waves\\ }

\end{center}

\begin{center}

\textbf{ By}

\end{center}

\begin{center}
\textbf{Muhammad Shoaib \\ e-mail: safridi@gmail.com}
\bigskip\bigskip\bigskip\bigskip\bigskip
\end{center}

\begin{center}
A thesis submitted in partial fulfilment \\
of the Requirements of Quaid-I-Azam University Islamabad, Pakistan \\
For the degree of Master of Philosophy

\end{center}

September 1999

\pagebreak

\begin{center}
\emph{I would like to dedicate this thesis to\\
 \textbf{my parents}\\}
\end{center}

\renewcommand{\contentsname}{Table of Contents}
\tableofcontents \listoffigures

\chapter{ GRAVITATIONAL WAVES}

\section{ Introduction}

Almost every student of science is familiar with the phenomenon of waves,
such as water waves ( ripples rolling across the ocean), sound waves
(vibration in the air), etc. The above mentioned kinds of waves are easily
observable. Electromagnetic waves are comparatively difficult to understand
but not as difficult as gravitational waves. The existence of gravitational
waves was disputed for a long time but is generally accepted now.

What are gravitational waves? How do they propagate? and what is their
energy content? These questions are addressed in the first two chapters. In
the third chapter the pseudo-Newtonian formalism and its extension is
reviewed in general and the formula for the momentum imparted to test
particles in arbitrary spacetimes is reviewed in particular. In chapter four
the analysis of a paper claiming to determine the spin for gravitational
waves is given, and compared with the spin given by a geodesic analysis. It
is demonstrated that the other claim is inconsistent. Finally in chapter
five a summary of the work is given with the conclusion.

\section{ Idealization}

The only way to come to grips with so complicated a subject as
General Relativity is by idealization{ . }Study one idealization
after another. Build a catalogue of idealization, of their
properties and of techniques for analyzing them and then arrive at
a conclusion.

Let's now see how can we idealize gravitational waves. As one idealizes
``water waves'' as small ripples of geometry rolling across the ocean so one
gives the name ``gravitational waves'' to small ripples rolling across the
space time. Both of the waves are idealizations. One cannot, with infinite
accuracy, delineate at any moment which drops of water are in the waves and
which are in the underlying ocean. Similarly one cannot tell precisely which
parts of spacetime are in the ripples and which are in the cosmological
backgrounds. One can almost do so; otherwise one would not speak of
``waves''. Look at the ocean, the seascape is dominated by the waves.
Changes occur at the surface of the ocean, which propagates obeying the
following [1] wave equation:

\begin{equation}
(\frac{1}{g^{2}}\frac{\partial ^{4}}{\partial t^{4}}+\frac{\partial ^{2}}{%
\partial y^{2}}+\frac{\partial ^{2}}{\partial x^{2}})(\textrm{height of the
surface)}=0.  \label{1.1}
\end{equation}
Similarly gravitational waves are perturbations of spacetime. As I have
already stated that gravitational waves are ripples of geometry. So let me
support my statement. Take a massive body and disturb it violently, the near
field adjusts rapidly, but the far field must wait for the signal that the
mass has moved to propagate to it at a finite speed $``c"$. Thus there is a
travelling kink which falls off in strength with distance. Hence the
gravitational waves are small ripples rolling across the spacetime.

Now get more sophisticated. Notice from a space ship the large-scale
curvature of the ocean's surface-curvature because the Earth is round.
Curvature because the Earth, Sun and Moon pull the water. As waves propagate
long distances, this curvature bends their fronts and changes slightly their
simple wave equation. Spacetime is similar. Propagating through the
universe, according to Einstein's theory must be a complex pattern of
small-scale ripples in the spacetime curvature, ripples produced by binary
stars, by gravitational collapse, by explosion in galactic nuclei etc.

\section{ Linear approximation for the investigation of gravitational
waves}

We are discussing gravitational waves in the frame work of General
Relativity, a nonlinear field theory of gravity. Though many of the
interesting consequences of General Relativity comes from its non linearity
it is worthwhile to study its linear approximation. Linearization actually
leads one to the gravitational waves.

What makes General Relativity nonlinear? As the field in General Relativity
is the metric tensor, its appearance in the field equations non-linearly
gives rise to the non-linearity of General Relativity. We cannot change the
way the metric tensor enters into the curvature but we can write the curved
spacetime metric as the flat spacetime metric tensor, $\eta _{\mu \nu }$,
and an additional term, $h_{\mu \nu }.$ We then require that $h_{\mu \nu }$
and its derivative occur only once in the field equations and higher powers
be neglected[6].

\begin{equation}
g_{\mu \nu }=\eta _{\mu \nu }+h_{\mu \nu }.
\end{equation}

Let

\begin{equation}
g^{\mu \nu }=\eta ^{\mu \nu }+f^{\mu \nu },
\end{equation}
where $\eta ^{\mu \nu }\;$is the inverse of $\eta _{\mu \nu }$. Then by
definition,

\begin{equation}
g^{\mu \rho }g_{\rho \pi }=(\eta ^{\mu \rho }+f^{\mu \rho })(\eta _{\rho \pi
}+h_{\rho \pi }),
\end{equation}

Which implies that

\begin{equation}
\delta _{\pi }^{\mu }=\delta _{\pi }^{\mu }+f^{\mu \rho }\eta _{\rho \pi
}+\eta ^{\mu \rho }h_{\rho \pi }+f^{\mu \rho }h_{\rho \pi }.
\end{equation}
Cancelling $\delta _{\pi }^{\mu }$ on both sides and multiplying through by $%
\eta ^{\nu \pi }$ we get

\begin{equation}
f^{\mu \nu }+\eta ^{\mu \rho }\eta ^{\nu \pi }h_{\rho \pi }+\eta ^{\nu \pi
}f^{\mu \rho }h_{\rho \pi }=0.
\end{equation}
The last term is clearly quadratic in the difference between the curved and
flat spacetime metric tensor. Thus to first order,

\begin{equation}
f^{\mu \nu }=-\eta ^{\mu \rho }\eta ^{\nu \pi }h_{\rho \pi }+O(h^{2}).
\end{equation}
Using the flat spacetime metric tensor to raise and lower indices we can
write

\begin{equation}
f^{\mu \nu }=-h^{\mu \nu }+O(h^{2}).
\end{equation}
Using equation (1.3), equation (1.8) becomes

\begin{eqnarray}
g^{\mu \nu } &=&\eta ^{\mu \nu }-h^{\mu \nu }+O(h^{2}) \\
&\approx &\eta ^{\mu \nu }-h^{\mu \nu }.  \nonumber
\end{eqnarray}

Using this linearization and for the moment taking Cartesian coordinates, so
that there are no derivatives of $\eta ^{\mu \nu },\,$the Christoffel
symbols linearize to:

\begin{equation}
\left\{
\begin{array}{c}
\rho \\
\mu \;\upsilon
\end{array}
\right\} \approx \frac{1}{2}\eta ^{\rho \pi }(h_{\mu \nu ,\pi }+h_{\nu \pi
,\mu }-h_{\mu \nu ,\pi }).
\end{equation}
Clearly terms quadratic in the Christoffel symbols become quadratic in $h$
and can be neglected compared with linear terms. Thus the linearized Ricci
tensor is

\begin{eqnarray}
R_{\mu \nu } &=&\left\{
\begin{array}{c}
\rho \\
\mu \;\upsilon
\end{array}
\right\} _{,\rho }-\left\{
\begin{array}{c}
\rho \\
\mu \;\rho
\end{array}
\right\} _{,\nu } \\
&\approx &\frac{1}{2}[\eta ^{\rho \pi }(h_{\mu \pi ,\nu }+h_{\nu \pi ,\mu
}-h_{\mu \nu ,\pi })]_{,\rho } \\
&&-\frac{1}{2}[\eta ^{\rho \pi }(h_{\mu \pi ,\rho }+h_{\rho \pi ,\mu
}-h_{\mu \rho ,\pi })]_{,\nu } \\
&=&\frac{1}{2}\eta ^{\rho \pi }(h_{\mu \pi ,\nu \rho }+h_{\nu \pi ,\mu \rho
}-h_{\mu \nu ,\rho \pi }-h_{\rho \pi ,\mu \nu })
\end{eqnarray}

A choice of coordinates can be made to have the first two and the last term
in the brackets disappear. To see this first we note that we can rewrite the
Ricci tensor as:

\begin{eqnarray}
R_{\mu \nu } &\approx &\frac{1}{2}(h_{\mu }^{\rho }-\frac{1}{2}h\delta _{\mu
}^{\rho })_{,\rho \nu }+\frac{1}{2}(h_{\nu }^{\rho }-\frac{1}{2}h\delta
_{\nu }^{\rho })_{,\mu \rho }  \nonumber \\
&&\;\;\;\;\;\;\;\;\;\;\;\;\;\;\;\;\;\;\;\;\;\;\;\;\;-\frac{1}{2}\eta ^{\rho
\pi }h_{\mu \nu ,\rho \pi },
\end{eqnarray}
where $h=h_{\mu }^{\mu }$ then we can break the first bracket in to two
terms. Now consider an infinitesimal transformation

\begin{equation}
x^{\mu }\longrightarrow x^{^{\prime }\mu }=x^{\mu }+\xi ^{\mu }(x^{\rho }),
\end{equation}
so that the terms quadratic in $h$ or $\xi $ can be neglected. Since this is
only a coordinate transformation, it must leave the metric invariant and
hence

\begin{eqnarray}
ds^{2} &=&g_{\mu \nu }(x^{\rho })dx^{\mu }dx^{\nu }=g_{\mu \nu }^{^{\prime
}}(x^{\prime \rho })dx^{^{\prime }\mu }dx^{\prime \nu }  \nonumber \\
&=&g_{\mu \nu }^{\prime }(x^{\prime \rho })(dx^{\mu }+\xi _{,\alpha }^{\mu
}dx^{\alpha })(dx^{\nu }+\xi _{,\beta }^{\nu }dx^{\beta }).
\end{eqnarray}
Using the linearization procedure it is easy to see that

\begin{equation}
h_{\mu \nu }^{^{\prime }}\approx h_{\mu \nu }-\xi _{\mu ,\nu }-\xi _{\nu
,\mu }.
\end{equation}
Thus we have

\begin{eqnarray}
(h_{\mu }^{\rho }-\frac{1}{2}h\delta _{\mu }^{\rho }) &=&(h_{\mu }^{\prime
\rho }-\frac{1}{2}h^{\prime }\delta _{\mu }^{\rho })-\eta ^{\nu \rho }\xi
_{\mu ,\nu }  \nonumber \\
&&\,\,\,\,\,\,\,\,\,\,\,\,\,\,\,\,\,\,\,\,\,\,\,\,\,\,\,\,\,\,\,\,\,\,-\,\xi
_{,\nu }^{\rho }+\xi _{,\nu }^{\nu }\delta _{\mu }^{\rho }.
\end{eqnarray}

Differentiating the above equation relative to $x^{\rho }\,$it is easy to
see that the last two terms simply cancel and we have

\begin{equation}
(h_{\mu }^{\rho }-\frac{1}{2}h\delta _{\mu }^{\rho })_{,\rho }=(h_{\mu
}^{\prime \rho }-\frac{1}{2}h^{\prime }\delta _{\mu }^{\rho })_{,\rho }-\eta
^{\nu \rho }\xi _{\mu ,\nu \rho }.  \label{eq14}
\end{equation}
Taking the harmonic gauge condition
\begin{equation}
\square \xi _{\mu }=-\phi _{\mu ,\rho }^{\rho },
\end{equation}
we can make

\begin{equation}
(h_{\mu }^{\rho }-\frac{1}{2}h\delta _{\mu }^{\rho })_{,\rho }=\phi _{\mu
,\rho }^{\rho }=0.  \label{eq15}
\end{equation}

It should be pointed out here that the use of Cartesian coordinates was not
crucial, but merely for convenience. With any other coordinates we could
have to introduce the corresponding flat spacetime Christoffel symbols. This
has been avoided so as not cause confusion of notation. We now drop the
primes and so obtain
\begin{equation}
R_{\mu \nu }\approx -\frac{1}{2}\eta ^{\rho \pi }h_{\mu \nu ,\rho \pi }=-%
\frac{1}{2}\square h_{\mu \nu }.
\end{equation}

We can contract Equation (\ref{eq15}) to obtain the Ricci scalar ($R=-\frac{1%
}{2}\square h)$ and hence combine them to obtain the Einstein tensor. Thus
the Einstein field equation becomes
\begin{equation}
\square \phi _{\mu \nu }=-2\kappa T_{\mu \nu }.
\end{equation}
Here $T_{\mu \nu }$ is the stress energy tensor and $\kappa $ is the
proportionality constant. In regions where $T_{\mu \nu }=0,$ $\phi _{\mu \nu
}$ then satisfies the wave equation, Thus $\phi _{\mu \nu }$ represent
gravitational waves.

\section{ Plane wave solution in the linearized theory}

The simplest of all solutions to the linearized equation $h_{\mu \nu ,\alpha
}^{\alpha }=0$ is the monochromatic plane wave solution [1]

\begin{equation}
h_{\mu \nu }=\Re [A_{\mu \nu }\exp (ik_{\alpha }x^{\alpha })].
\end{equation}
Here $\Re $ means that one must take the real part in the bracket. While $%
A_{\mu \nu }\,$is the amplitude and $k_{\mu }$ is the wave vector, satisfying

\begin{eqnarray}
k_{\alpha }k^{\alpha } &=&0\,\,\,\,\,\,\,\,\,\,\,\,\,\,\textrm{(}{\bf k}\textrm{%
\thinspace a\thinspace null\thinspace vector),} \\
A_{\mu \alpha }k^{\alpha } &=&0\,\,\,\,\,\,\,\,\,\,\,\,({\bf
A}\textrm{ orthogonal to }{\bf k}\textrm{).}  \nonumber
\end{eqnarray}
This solution describes a wave with the frequency

\[
\omega =k^{0}=(k_{x}^{2}+k_{y}^{2}+k_{z}^{2})^{1/2},
\]
which propagate with the speed of light in the direction $%
\,(1/k^{0})(k_{x},k_{y},k_{z})$. At first sight the amplitude
$A_{\mu \nu \textrm{ }}$ appears to have six independent
components (ten, less the four orthogonality constraints $A_{\mu
\alpha }k^{\alpha }=0$). But this cannot be right. The
gravitational field has two dynamic degrees of freedom (not six).

The plane wave vector

\begin{equation}
\xi ^{\mu }=-iC^{\mu }\exp (ik_{\alpha }x^{\alpha }),
\end{equation}
with four arbitrary constants $C^{\mu },$ generates a gauge transformation
that can change arbitrarily four of the six independent components of $%
A_{\mu \nu }.\,$One gets rid of this arbitrariness by choosing a specific
gauge.

\section{ The transverse traceless (TT) gauge}

Consider the four velocity ${\bf u}$ through all the space time and impose
the condition [1]

\begin{equation}
A_{\mu \nu }u^{\nu }=0.
\end{equation}
These are only three constraints on $A_{\mu \nu }$. Why not four? Because $%
{\bf A}$ is orthogonal to ${\bf \,k}$ i.e. $k^{\mu }(A_{\mu \nu }u^{\nu })=0$%
, as already mentioned. As a fourth constraint use a gauge transformation to
set$\,\,A_{\mu }^{\mu }=0.$ So there are now four constraints in all, $%
A_{\mu \alpha }u^{\alpha }=A_{\mu \alpha }k^{\alpha }=A_{\alpha }^{\alpha
}=0,$ on the ten components of amplitude. Thus the two remaining free
components of $A_{\mu \nu }$ represents the two degrees of freedom in the
plane gravitational wave.

It is useful to restate the four constraints,$\,A_{\mu \alpha }u^{\alpha
}=A_{\mu \alpha }k^{\alpha }=A_{\alpha }^{\alpha }=0$ in the Lorentz frame
where $u^{0}=1,u^{j}=0$ and in a frame where $k^{\alpha }$ does not appear
explicitly:

\begin{equation}
h_{\mu 0}=0\textrm{ , i.e only the spatial componemts
}h_{jk\textrm{ }}\textrm{are non-zero}  \label{1}
\end{equation}
\begin{equation}
h_{kj,j}=0\textrm{ i.e the spatial components are divergence free}
\label{2}
\end{equation}
\begin{equation}
h_{kk}=h_{\mu }^{\mu }=0\textrm{ i.e the spatial components are
trace free.} \label{3}
\end{equation}
The gauge conditions are all linear in $h_{\mu \nu }$ so an arbitrary wave
will also satisfy the above gauge conditions. In this gauge only the $h_{jk}$
are non-zero. So we need only to impose the six wave equations
\begin{equation}
\square h_{jk}=h_{jk,\alpha \alpha }=0
\end{equation}
Now the definition. Any symmetric tensor satisfying the constraints (\ref{1}%
), (\ref{2}) and (\ref{3}) is called a transverse traceless tensor. Why it
is called transverse traceless? Because it is purely spatial ($h_{\mu 0}=0$)
and, if thought of as a wave is transverse to the direction of propagation ($%
h_{ij,j}=h_{ij}k_{j}=0$) and traceless because $h_{kk}=0$.

The special gauge in which $h_{\mu \nu }$ reduces to its transverse
traceless part is called the transverse traceless gauge . The conditions (%
\ref{1}), (\ref{2}) and (\ref{3}) defining this gauge can be summarized as:
\begin{equation}
h_{\mu \nu }=h_{\mu \nu }^{TT}.
\end{equation}

Only pure waves can be reduced to TT gauge. In the TT-gauge the space
components
\begin{equation}
R_{j0ko}=R_{0j0k}=-R_{j00k}=-R_{0jk0}.
\end{equation}
of the Riemannian curvature tensor have an especially simple form
\begin{equation}
R_{j0k0}=-\frac{1}{2}h_{jk,00}^{TT}.
\end{equation}
As the curvature tensor is gauge invariant therefore $h_{\mu \nu }$ cannot
be reduced to still fewer components than it has in the TT-gauge.

\section{ Comparison with electromagnetic waves}

A simple system consisting of two charges of equal magnitude but of opposite
signs, each situated at a distance $r/2$ from the origin $O$ (see $Fig.1.1$%
), which is taken to lie on the line connecting the charges, is
the simplest example of an electric dipole. The oscillation in the
dipole generates electromagnetic waves.\smallskip

\begin{figure}
\centerline{\epsfig{file=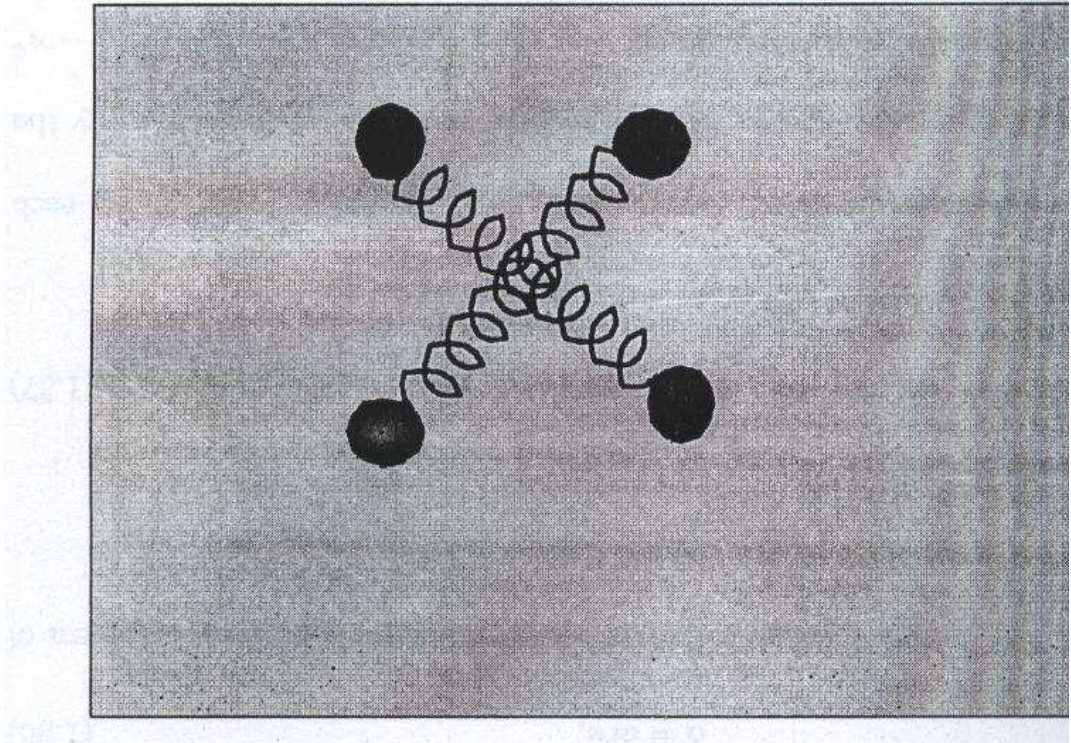,height=10cm,angle = 0}}
\caption{A system consisting of four masses attached by springs,
forming two dipoles. The oscillation in the system generate
gravitational waves.}
 \label{}
\end{figure}

Now we define the electric dipole moment of the pair of equal charges as the
product of charge $q$ and the separation $r,$

\begin{equation}
{\bf d}=qr{\bf e,}  \label{3.30}
\end{equation}
where ${\bf e}$ is the unit vector from negative to positive charge.
Consider a system of charges $q_{a}$ and let $r_{a}$ be their radius vectors
then

\begin{equation}
{\bf d}={\sum_a }q_{a}r_{a},  \label{3.31}
\end{equation}
is called the dipole moment of the system of charges.

Consider masses in an isolated, nearly Newtonian system, moving about each
other. They emit radiations. For an order of magnitude estimate, one can
apply the familiar radiation formula of electromagnetic theory, with the
replacement $q^{2}\rightarrow -m^{2}$ which converts the static Coulomb's
law into Newton's law of attraction. Although it introduces a moderate error
in numerical factor and changes angular distribution, but it gives an
estimate of the total power radiated. In electromagnetic theory, electric
dipole radiation dominates, with power output or luminosity ``$L$'' given
by[1],
\begin{equation}
{\bf L}_{elec\,\,dip}=\frac{2}{3}q^{2}a^{2},
\end{equation}
where $a$ is the acceleration in the dipole. Then using equation (\ref{3.30}%
) we can write

\begin{equation}
{\bf L}_{elec\,dip}=\frac{2}{3}\stackrel{\cdot \cdot }{d}^{2}.  \label{3.33}
\end{equation}
The gravitational analogue of the electric dipole moment is the mass dipole
moment,

\begin{equation}
{\bf d\,}={\sum_{\,A} }m_{A}x_{A},  \label{3.34}
\end{equation}
its first time-rate of change is the total momentum of the system,

\begin{equation}
\stackrel{.}{{\bf d\,}}=\sum m_{A}\stackrel{.}{x_{A}^{.}=\,{\bf P}},
\end{equation}
where ${\bf P}$ is the momentum of the system. The time-rate of change of
the mass dipole moment has to vanish because of the law of conservation of
momentum, $\stackrel{..}{d\,\,}=\,\stackrel{.}{P\,}=\,0$ therefore there can
be no mass dipole radiation in gravitation physics.

The next strongest type of electromagnetic radiation are magnetic dipole and
electric quadrupole. Magnetic dipole radiation is generated by the second
time derivative of the magnetic moment,$\,\stackrel{..}{\mu }.$ Here again
the gravitational analogue is a constant of motion of the angular momentum
i.e.,

\begin{eqnarray}
\mu &=&\sum (position\,of\,\,A)\times (current\,due\,to\,A)  \nonumber \\
&=&\sum r_{A}\times (mV_{A})={\bf J}=\textrm{{\it constant}}
\end{eqnarray}
Which shows that $\stackrel{..}{\mu }=0,$ so it can not radiate. Thus, there
can be no gravitational dipole radiation of any sort. Physical example is a
system of two masses attached by a spring, acts as an oscillating dipole.

\subsection{ Comparison with plane electromagnetic waves}

Consider the metric [1] (with signature $(+,-,-,-)$)

\begin{equation}
ds^{2}=L^{2}(u)(dx^{2}+dy^{2})-dudv.  \label{A}
\end{equation}
where $u=t-z\,$\thinspace and$\,v=t+z\,\,\,$which is always flat. It
satisfies the vacuum Einstein equations $R_{\mu \nu }=0\,\,$which implies
that$\,\,L^{\prime \prime }=0\,\;$(see Appendix 1 for the proof). Now if $%
L^{\prime \prime }=0$ the spacetime is static therefore cannot represent
gravitational waves. In this metric the electromagnetic potential\thinspace
\thinspace \thinspace \thinspace \thinspace \thinspace \thinspace
\begin{equation}
{\bf A}=A_{\mu }{\bf d}x^{\mu }=A(u)dx  \label{B}
\end{equation}
satisfies the Maxwell equation for arbitrary $A(u).$ It represents an
electromagnetic plane wave analogous to the gravitational plane wave. The
only non-zero components of this wave are

\begin{equation}
F_{ux}=A^{\prime }\;i.e\;\,\,\,\,\,\,\,\,\,F_{tx}=-F_{zx}=A^{\prime }.\,\,
\end{equation}
$\,\,\,\,\,\,$

So the wave propagates in the $z$-direction, the magnetic vector oscillates
in the $y$-direction and electric vector in the $x$-direction. The only
non-zero vector of the stress energy tensor is
\begin{equation}
T_{uu}=(4\pi L^{2})^{-1}(A^{\prime })^{2}
\end{equation}
It can easily be verified that the Maxwell equation are satisfied by (\ref{B}%
) in (\ref{A}).

To make the metric acceptable, we need to impose the Einstein equations $%
R_{\mu \nu }-\frac{1}{2}g_{\mu \nu }R=\kappa T_{\mu \nu },\kappa $ is
generally $\frac{8\pi G}{c^{4}},$but here we use gravitational units i.e. $%
G=c=1$ so it becomes $8\pi .$ As all the nonvanishing Ricci tensor
components are the same and they read $\frac{-2L^{\prime \prime }}{L},$ so
the Einstein field equations becomes $L^{\prime \prime }+(4\pi T_{\mu \mu
})L=0.$ This has exactly the form of the equation $L^{\prime \prime }+(\beta
^{\prime })^{2}L=0,$ (which will be discussed in detail in chapter 2) for
gravitational plane wave.

\chapter{REALITY OF GRAVITATIONAL WAVES}

In the linearization procedure the stress energy tensor was taken to be
zero, which creates a conceptual problem, namely the question of reality of
gravitational waves. This will be discussed in detail in the first section
of this chapter. For this purpose some exact gravitational wave solutions of
the Einstein field equations are presented. Returning to the purpose of this
chapter the energy contents of the waves are discussed. Finally as an
additional but necessary, topic some sources of gravitational waves are
discussed with in the scope of this work.

\section{ The conceptual problem}

The exact gravitational waves being by definition, solutions of the vacuum
field equations, have a zero stress energy tensor. This creates a conceptual
problem. How can these solutions of the Einstein field equations represent
waves if they carry no energy? Essentially, this problem arises because
energy is not a well defined concept in General Relativity. For energy to be
well defined in General Relativity the metric must have a time like isometry
(Killing vector) so as to allow time translational invariance. This will not
generally be true. Infact, spacetimes for which it is true are static
whereas gravitational wave solutions must be non-static. Thus energy is not
well defined for spacetimes containing gravitational waves. The question
then is, how can we tell that these solutions really do behave as we would
expect of the waves?

One way to answer this question: the very process of linearization provides
the energy. The point is that if $\Box \phi _{\mu \nu }=0$ exactly then $%
R_{\mu \nu }\approx 0$ to order $h$ but $R_{\mu \nu }\neq 0$ to higher
orders in $h$. Thus $T_{\mu \nu }\approx 0$ only to order $h$ but $T_{\mu
\nu }\neq 0$ generally. Conversely if $T_{\mu \nu }=0\,$and\thinspace
\thinspace \thinspace hence$\,R_{\mu \nu }=0$ exactly then $\Box \phi _{\mu
\nu }\approx 0$ only to order $h$. Thus we can expand $R_{\mu \nu }$ in
powers of $h,\,$retaining linear terms on the left side of the equation and
transposing all higher powers to the right side. These higher order terms
become an effective stress energy tensor and the linearized equations give
the gravitational waves.

\section{ Some exact solutions of gravitational waves}

So far we have obtained the wave equations for gravity by linearizing the
Einstein field equations. In principle the solutions so obtained could be
exact solutions of the vacuum Einstein field equations, of course there
could be trivial static solutions which effectively satisfy the Laplace
equation. But they do not represent moving waves so we are interested in the
solutions which are non-static.

The first solution to be discovered was for cylindrical gravitational waves,
by Einstein and Rosen in (1937) [2]. So first consider this:

\subsection{ Cylindrical gravitational wave solution}

A cylindrically symmetric metric depending on two arbitrary functions $%
\gamma $ and $\psi $ of the time $t$ and cylindrical radial coordinate $\rho
$, is

\begin{equation}
ds^{2}=e^{2(\gamma -\psi )}(dt^{2}-d\rho ^{2})-e^{-2\psi }\rho ^{2}d\varphi
^{2}-e^{2\psi }dz^{2}.  \label{2.1}
\end{equation}

The metric tensor is

\begin{equation}
g_{ab}=\left(
\begin{array}{cccc}
e^{2(\gamma -\psi )} & 0 & 0 & 0 \\
0 & -e^{2(\gamma -\psi )} & 0 & 0 \\
0 & 0 & -\rho ^{2}e^{-2\psi } & 0 \\
0 & 0 & 0 & -e^{-2\psi }
\end{array}
\right) .
\end{equation}

Its inverse is

\begin{equation}
g^{ab}=\left(
\begin{array}{cccc}
e^{-2(\gamma -\psi )} & 0 & 0 & 0 \\
0 & -e^{-2(\gamma -\psi )} & 0 & 0 \\
0 & 0 & -\rho ^{-2}e^{-2\psi } & 0 \\
0 & 0 & 0 & -e^{2\psi }
\end{array}
\right)
\end{equation}

The non-zero Christoffel symbols are:

\[
\left\{
\begin{array}{c}
0 \\
0\,\,\,0
\end{array}
\right\} =\left\{
\begin{array}{c}
0 \\
1\,\,\,1
\end{array}
\right\} =\left\{
\begin{array}{c}
1 \\
1\,\,\,0
\end{array}
\right\} =\left\{
\begin{array}{c}
1 \\
0\,\,\,1
\end{array}
\right\} =\gamma ^{.}-\psi ^{.};
\]

\begin{equation}
\left\{
\begin{array}{c}
0 \\
0\,\,\,1
\end{array}
\right\} =\left\{
\begin{array}{c}
0 \\
1\,\,\,0
\end{array}
\right\} =\left\{
\begin{array}{c}
1 \\
0\,\,\,0
\end{array}
\right\} =\left\{
\begin{array}{c}
1 \\
1\,\,\,1
\end{array}
\right\} =\gamma ^{\prime }-\psi ^{\prime };
\end{equation}

\begin{equation}
\left\{
\begin{array}{c}
0 \\
2\,\,\,2
\end{array}
\right\} =-\rho ^{2}\psi ^{\cdot }e^{-2\gamma
};\,\,\,\,\,\,\,\,\,\,\,\,\left\{
\begin{array}{c}
0 \\
3\,\,\,3
\end{array}
\right\} =\psi ^{\cdot }e^{2(2\psi -\gamma
)};\,\,\,\,\,\,\,\,\,\,\,\,\,\,\,\,\,\,\,\,\,
\end{equation}

\begin{equation}
\left\{
\begin{array}{c}
1 \\
2\,\,\,2
\end{array}
\right\} =\rho (\rho \psi ^{\prime }-1)e^{-2\gamma };\,\,\,\,\,\,\,\left\{
\begin{array}{c}
1 \\
3\,\,\,3
\end{array}
\right\} =-\psi ^{\prime }e^{2(2\psi -\gamma )}\,;\,\,\,\,\,\,\,\,\,
\end{equation}

\begin{equation}
\,\,\,\left\{
\begin{array}{c}
2 \\
0\,\,\,2
\end{array}
\right\} =\,\,\,\left\{
\begin{array}{c}
2 \\
2\,\,\,0
\end{array}
\right\} =-\psi ^{\cdot };\left\{
\begin{array}{c}
2 \\
2\,\,\,1
\end{array}
\right\} =-(\psi ^{\prime }-1/\rho )\,;\,\,\,\,\,\,
\end{equation}

\begin{equation}
\left\{
\begin{array}{c}
3 \\
0\,\,3
\end{array}
\right\} =\left\{
\begin{array}{c}
3 \\
3\,\,\,0
\end{array}
\right\} =\psi ^{\cdot };\,\,\,\,\,\,\,\,\left\{
\begin{array}{c}
3 \\
1\,\,\,\,3
\end{array}
\right\} =\psi ^{\prime
}\,;\,\,\,\,\,\,\,\,\,\,\,\,\,\,\,\,\,\,\,\,\,\,\,\,\,\,\,\,\,\,\,\,\,\,\,\,
\end{equation}

\begin{equation}
\left\{
\begin{array}{c}
\mu \\
\mu \,\,0
\end{array}
\right\} =(\ln \sqrt{g})_{,0}=2(\gamma ^{.}-\psi
^{.})\,;\,\,\,\,\,\,\,\,\,\,\,\,\,\,\,\,\,\,\,\,\,\,\,\,\,\,\,\,\,\,\,\,\,\,%
\,\,\,\,\,\,\,\,\,\,\,\,\,\,\,\,\,\,\,\,\,\,\,\,\,\,\,\,\,\,\,
\end{equation}

\begin{equation}
\left\{
\begin{array}{c}
\mu \\
\mu \,\,1
\end{array}
\right\} =(\ln \sqrt{g})_{,1}=2(\gamma ^{^{\prime }}-\psi ^{^{\prime
}})+1/\rho
\,.\,\,\,\,\,\,\,\,\,\,\,\,\,\,\,\,\,\,\,\,\,\,\,\,\,\,\,\,\,\,\,\,\,\,\,\,%
\,\,\,\,\,\,\,\,\,
\end{equation}
The ``$\,^{\prime }\,$'' refers to differentiation with respect to $\rho $
and the dot differentiation with respect $t.\,$The non vanishing components
of the Ricci tensor are $R_{00},\,R_{11},\,R_{22}$ and $R_{33}$.\thinspace
Only four components are needed for the purpose. Using the Christoffel
symbols listed above we obtain the following Ricci tensor components,

\begin{eqnarray}
R_{00} &=&\left\{
\begin{array}{c}
\mu \\
0\,\,\mu
\end{array}
\right\} _{,\mu }-(\ln \sqrt{g})_{,00}+(\ln \sqrt{g})_{,\mu }\left\{
\begin{array}{c}
\mu \\
0\,\,\,\,\,0
\end{array}
\right\} -\left\{
\begin{array}{c}
\mu \\
0\,\,\,\,\,\nu
\end{array}
\right\} \left\{
\begin{array}{c}
\nu \\
0\,\,\,\,\,\,\mu \,\,
\end{array}
\right\} \,\,\,\,\,\,\,  \nonumber \\
&=&\left\{
\begin{array}{c}
0 \\
0\,\,0
\end{array}
\right\} _{,0}+\left\{
\begin{array}{c}
1 \\
0\,\,0
\end{array}
\right\} _{,1}-(\ln \sqrt{g})_{,00}+(\ln \sqrt{g})_{,0}\left\{
\begin{array}{c}
0 \\
0\,\,\,\,\,0
\end{array}
\right\}  \nonumber \\
&&-\left\{
\begin{array}{c}
0 \\
0\,\,\,\,\,0
\end{array}
\right\} +(\ln \sqrt{g})_{,1}\,\left\{
\begin{array}{c}
1 \\
0\,\,\,\,\,\,\,0
\end{array}
\right\} \,-2\left\{
\begin{array}{c}
0 \\
0\,\,\,\,\,1\,
\end{array}
\right\} \left\{
\begin{array}{c}
1 \\
0\,\,\,\,\,\,0\,\,
\end{array}
\right\}  \nonumber \\
&&-\left\{
\begin{array}{c}
2 \\
0\,\,\,\,\,\,2\,\,
\end{array}
\right\} ^{2}-\left\{
\begin{array}{c}
3 \\
0\,\,\,\,\,\,3\,\,
\end{array}
\right\} ^{2}
\end{eqnarray}
$\,\,\,\,\,\,\,\,\,\,\,\,\,\,$

$\,\,\,\,\,\,\,\,\,\,\,\,$%
\begin{eqnarray}
\,\,\,\,\,\,\,\,\,\,\,\, &=&\left\{
\begin{array}{c}
0 \\
0\,\,0
\end{array}
\right\} _{,0}+\left\{
\begin{array}{c}
1 \\
0\,\,0
\end{array}
\right\} _{,1}-(\ln \sqrt{g})_{,00}+(\ln \sqrt{g})_{,0}\left\{
\begin{array}{c}
0 \\
0\,\,\,\,\,0
\end{array}
\right\} -\left\{
\begin{array}{c}
0 \\
0\,\,\,\,\,0
\end{array}
\right\}  \nonumber \\
&&+(\ln \sqrt{g})_{,1}\,\left\{
\begin{array}{c}
1 \\
0\,\,\,\,\,\,\,0
\end{array}
\right\} \,-2\left\{
\begin{array}{c}
0 \\
0\,\,\,\,\,1\,
\end{array}
\right\} \left\{
\begin{array}{c}
1 \\
0\,\,\,\,\,\,0\,\,
\end{array}
\right\} -\left\{
\begin{array}{c}
2 \\
0\,\,\,\,\,\,2\,\,
\end{array}
\right\} ^{2}-\left\{
\begin{array}{c}
3 \\
0\,\,\,\,\,\,3\,\,
\end{array}
\right\} ^{2}\,\,\,\,
\end{eqnarray}
$\,\,\,\,\,\,\,\,\,\,\,\,\,\,\,\,\,\,\,\,\,\,\,\,\,\,\,\,\,\,\,\,\,\,\,\,\,%
\,\,\,\,\,\,\,\,\,\,\,\,\,\,\,\,\,\,\,\,\,\,\,\,\,\,\,\,\,\,\,\,\,\,\,\,\,\,%
\,\,\,\,\,\,\,\,\,\,\,\,\,\,\,\,\,\,\,\,\,$

After simplification we get

\begin{eqnarray}
R_{00} &=&\gamma ^{\cdot \cdot }-\psi ^{\cdot \cdot }+\gamma ^{\prime \prime
}-\psi ^{\prime \prime }-2(\gamma ^{\cdot \cdot }-\psi ^{\cdot \cdot
})+(\gamma ^{\cdot }-\psi ^{\cdot })^{2}-2(\gamma ^{\prime }-\psi ^{\prime
})^{2}  \nonumber \\
&&-(\gamma ^{\cdot }-\psi ^{\cdot })^{2}-2(\psi ^{\cdot })^{2}+[\frac{1}{%
\rho }+2(\gamma ^{\prime }-\psi ^{\prime })](\gamma ^{\prime }-\psi ^{\prime
})\,\,\,\,\,\,\,\,\,\,\,\,\,\,\,\,\,\,\,  \nonumber \\
&=&-(\gamma ^{\cdot \cdot }-\psi ^{\cdot \cdot })+(\gamma ^{\prime \prime
}-\psi ^{\prime \prime })+\frac{1}{\rho }(\gamma ^{\prime }-\psi ^{\prime
})-2(\psi ^{\cdot
})^{2}.\,\,\,\,\,\,\,\,\,\,\,\,\,\,\,\,\,\,\,\,\,\,\,\,\,\,\,\,\,\,\,\,\,\,%
\,\,\,\,
\end{eqnarray}
Similarly other non-vanishing components are,

\begin{equation}
R_{11}=(\gamma ^{\cdot \cdot }-\psi ^{\cdot \cdot })-(\gamma ^{\prime \prime
}-\psi ^{\prime \prime })+\frac{1}{\rho }(\gamma ^{\prime }+\psi ^{\prime
})-2(\psi ^{\prime \prime
})^{2}.\,\,\,\,\,\,\,\,\,\,\,\,\,\,\,\,\,\,\,\,\,\,\,\,\,\,\,\,\,\,\,\,\,\,%
\,\,\,\,\,\,\,\,
\end{equation}

\begin{equation}
R_{22}=\rho ^{2}e^{-\gamma }(-\psi ^{\cdot \cdot }+\psi ^{\prime \prime }+%
\frac{1}{\rho }\psi ^{\prime
})\,.\,\,\,\,\,\,\,\,\,\,\,\,\,\,\,\,\,\,\,\,\,\,\,\,\,\,\,\,\,\,\,\,\,\,\,%
\,\,\,\,\,\,\,\,\,\,\,\,\,\,\,\,\,\,\,\,\,\,\,\,\,\,\,\,\,\,\,\,\,\,\,\,\,\,%
\,\,\,\,\,\,\,\,\,\,\,\,\,\,\,\,\,\,\,\,\,\,\,\,\,\,\,\,\,\,\,\,\,\,\,
\end{equation}

\begin{equation}
R_{01}=\frac{1}{\rho }\gamma ^{\cdot }-2\psi ^{\cdot }\psi ^{\prime
}.\,\,\,\,\,\,\,\,\,\,\,\,\,\,\,\,\,\,\,\,\,\,\,\,\,\,\,\,\,\,\,\,\,\,\,\,\,%
\,\,\,\,\,\,\,\,\,\,\,\,\,\,\,\,\,\,\,\,\,\,\,\,\,\,\,\,\,\,\,\,\,\,\,\,\,\,%
\,\,\,\,\,\,\,\,\,\,\,\,\,\,\,\,\,\,\,\,\,\,\,\,\,\,\,\,\,\,\,\,\,\,\,\,\,\,%
\,\,\,\,\,\,\,\,\,\,\,\,\,\,\,\,\,\,\,\,\,\,\,\,\,\,\,\,\,\,\,\,
\end{equation}
So the Einstein field equations becomes

\begin{equation}
(\gamma ^{\cdot \cdot }-\psi ^{\cdot \cdot })-(\gamma ^{\prime \prime }-\psi
^{\prime \prime })-\frac{1}{\rho }(\gamma ^{\prime }-\psi ^{\prime })+2(\psi
^{\cdot
})^{2}=0.\,\,\,\,\,\,\,\,\,\,\,\,\,\,\,\,\,\,\,\,\,\,\,\,\,\,\,\,\,\,\,\,\,%
\,\,\,\,\,\,\,\,\,\,\,\,\,\,\,\,\,\,\,\,\,\,\,\,\,  \label{1a}
\end{equation}
$\,\,\,$

\begin{equation}
(\gamma ^{..}-\psi ^{..})-(\gamma ^{\prime \prime }-\psi ^{\prime \prime })+%
\frac{1}{\rho }(\gamma ^{\prime }+\psi ^{\prime })-2(\psi ^{\prime \prime
})^{2}=0.\,\,\,\,\,\,\,\,\,\,\,\,\,\,\,\,\,\,\,\,\,\,\,\,\,\,\,\,\,\,\,\,\,%
\,\,\,\,\,\,\,\,\,\,\,\,\,\,\,\,\,\,\,\,\,\,  \label{2a}
\end{equation}

\begin{equation}
\rho ^{2}e^{-\gamma }(-\psi ^{\cdot \cdot }+\psi ^{\prime \prime }+\frac{1}{%
\rho }\psi ^{\prime
})\,=0.\,\,\,\,\,\,\,\,\,\,\,\,\,\,\,\,\,\,\,\,\,\,\,\,\,\,\,\,\,\,\,\,\,\,%
\,\,\,\,\,\,\,\,\,\,\,\,\,\,\,\,\,\,\,\,\,\,\,\,\,\,\,\,\,\,\,\,\,\,\,\,\,\,%
\,\,\,\,\,\,\,\,\,\,\,\,\,\,\,\,\,\,\,\,\,\,\,\,\,\,\,\,\,\,\,\,\,\,\,\,\,\,%
\,\,\,\,\,\,\,\,\,\,\,\,\,  \label{a}
\end{equation}

\begin{equation}
\frac{1}{\rho }\gamma ^{\cdot }-2\psi ^{\cdot }\psi ^{\prime
}\,=0.\,\,\,\,\,\,\,\,\,\,\,\,\,\,\,\,\,\,\,\,\,\,\,\,\,\,\,\,\,\,\,\,\,\,\,%
\,\,\,\,\,\,\,\,\,\,\,\,\,\,\,\,\,\,\,\,\,\,\,\,\,\,\,\,\,\,\,\,\,\,\,\,\,\,%
\,\,\,\,\,\,\,\,\,\,\,\,\,\,\,\,\,\,\,\,\,\,\,\,\,\,\,\,\,\,\,\,\,\,\,\,\,\,%
\,\,\,\,\,\,\,\,\,\,\,\,\,\,\,\,\,\,\,\,\,\,\,\,\,\,\,\,\,\,\,\,\,\,\,\,\,\,%
\,\,\,\,\,\,\,\,\,  \label{3a}
\end{equation}

Equation (\ref{a}) is the usual cylindrical form of the wave$\,.$ As this is
a second order linear differential equation, the solution has two arbitrary
constants, one corresponding to the ingoing cylindrical wave and the other
corresponding to the outgoing. Retaining only the outgoing waves with
amplitude $A$ and frequency $\omega $ we have

\begin{equation}
\psi (t,\rho )=A[J_{0}(x)\cos (\omega t)+N_{0}(x)\sin (\omega t)].  \label{4}
\end{equation}
where $x=\omega \rho \,,\,J_{0}(x)\,$\thinspace and$\,\,N_{0}(x)\,$ are the
zero order Bessel and Neuman functions respectively. Add equation (\ref{1a})
and (\ref{2a}) to obtain

\begin{equation}
\gamma ^{\prime }=\rho (\psi ^{\cdot 2}+\psi ^{\prime 2})  \label{5}
\end{equation}

Equations (\ref{3a}) and (\ref{5}) gives the space and time derivative of $%
\gamma (t,\rho )$ in terms of functions that are now known through equation (%
\ref{4}). The only thing required is the integration with respect to each
variable. The time integration is relatively easy and the space integration,
though tedious, is in principle easy (using the standard formulae for the
integrals of the Bessel and Neumann function). The resulting solution of $%
\gamma $ is then

\begin{equation}
\left.
\begin{array}{c}
\gamma (t,\rho )=\frac{1}{2}A^{2}x\{J_{0}(x)J_{0}^{\prime
}(x)+N_{0}(x)N_{0}^{\prime }(x)+x[J_{0}^{\prime }(x)^{2}+N_{0}^{\prime
}(x)^{2}] \\
+[J_{0}(x)J_{0}^{\prime }(x)-N_{0}(x)N_{0}^{\prime }(x)]\cos (2\omega t) \\
\,\,\,\,\,\,\,\,\,\,\,\,\,\,\,\,\,\,\,\,\,\,\,\,\,\,\,\,\,\,\,\,\,\,%
\,+[J_{0}(x)N_{0}^{\prime }(x)+J_{0}^{\prime }(x)N_{0}(x)]\sin (2\omega t)\}-%
\frac{2}{\pi }A^{2}\omega t
\end{array}
\right\} .  \label{6*}
\end{equation}
Where prime now refers to differentiation with respect to $x$ not $\rho $.
Hence the metric

\begin{equation}
ds^{2}=e^{2(\gamma -\psi )}(dt^{2}-d\rho ^{2})-e^{-2\psi }\rho ^{2}d\varphi
^{2}-e^{2\psi }dz^{2}.
\end{equation}
represents cylindrical gravitational wave with the above definition of $%
\gamma \,$and $\psi \,.$

\subsection{ Plane gravitational wave solution}

The solution which we are going to present here was discovered by Bondi and
Robinson in 1957 [3]. We take a line element which incorporates the
symmetries of a plane and represents a wave going in the x-direction.

\begin{equation}
ds^{2}=e^{2\Omega (u)}(dt^{2}-dx^{2})-u^{2}(e^{2\beta (u)}dy^{2}-e^{-2\beta
(u)}dz^{2}).  \label{7}
\end{equation}
$\,\,\,\,$Here all coefficients in the metric are functions of $(t-x)$ which
are represented by $u.$ The$\,(x,y,z)$ are not the usual Cartesian
coordinates but rectangular coordinates in a curved space-time. Thus the
metric tensor of equation(\ref{7}) is

\begin{equation}
g_{ab}=\left(
\begin{array}{cccc}
e^{2\Omega (u)} & 0 & 0 & 0 \\
0 & -e^{2\Omega (u)} & 0 & 0 \\
0 & 0 & -u^{2}e^{2\beta (u)} & 0 \\
0 & 0 & 0 & -u^{2}e^{-2\beta (u)}
\end{array}
\right) .
\end{equation}
Its inverse can be written as

\begin{equation}
g^{ab}=\left(
\begin{array}{cccc}
e^{-2\Omega (u)} & 0 & 0 & 0 \\
0 & -e^{-2\Omega (u)} & 0 & 0 \\
0 & 0 & -u^{-2}e^{-2\beta (u)} & 0 \\
0 & 0 & 0 & -u^{-2}e^{2\beta (u)}
\end{array}
\right) .
\end{equation}

Here it should be noted for convenience that

$\Omega _{,0}=\Omega ^{\prime }=-\Omega _{,1}$\thinspace and$\,\,\beta
_{,0}=\beta ^{\prime }=-\beta _{,1}.$ Here `` $\prime $ '' represents
derivative with respect to $u$.$\,$%
\begin{eqnarray}
&&^{\left\{
\begin{array}{c}
0 \\
0\,\,\,0
\end{array}
\right\} =\left\{
\begin{array}{c}
0 \\
1\,\,\,1
\end{array}
\right\} =\left\{
\begin{array}{c}
1 \\
1\,\,\,0
\end{array}
\right\} =\left\{
\begin{array}{c}
1 \\
0\,\,\,1
\end{array}
\right\} =-\left\{
\begin{array}{c}
0 \\
0\,\,\,1
\end{array}
\right\} =-\left\{
\begin{array}{c}
1 \\
0\,\,\,0
\end{array}
\right\} }\,\,\,\,\,\,\,\,\,\,\,\,\,\, \\
&=&-\left\{
\begin{array}{c}
1 \\
1\,\,\,1
\end{array}
\right\} =-\Omega ^{\prime };  \nonumber
\end{eqnarray}

\begin{eqnarray}
\left\{
\begin{array}{c}
0 \\
2\,\,\,2
\end{array}
\right\} &=&\left\{
\begin{array}{c}
1 \\
2\,\,\,2
\end{array}
\right\} =u(u\beta ^{\prime }+1)e^{2(\beta -\Omega
)}\,\,\,;\,\,\,\,\,\,\,\,\,\,\,\,\,\,\,\,\,\,\,\,\,\,\,\,\,\,\,\,\,\,\,\,\,%
\,\,\,\, \\
\left\{
\begin{array}{c}
0 \\
3\,\,\,3
\end{array}
\right\} &=&\left\{
\begin{array}{c}
1 \\
3\,\,\,3
\end{array}
\right\} =u(-u\beta ^{\prime }+1)e^{-2(\beta +\Omega )};  \nonumber
\end{eqnarray}

\begin{equation}
\,\,\,\,\,\,\,\,\,\,\,\,\,\left\{
\begin{array}{c}
2 \\
0\,\,\,2
\end{array}
\right\} =\,\,\,\left\{
\begin{array}{c}
2 \\
2\,\,\,0
\end{array}
\right\} =-\left\{
\begin{array}{c}
2 \\
2\,\,\,1
\end{array}
\right\} =\left\{
\begin{array}{c}
2 \\
2\,\,\,1
\end{array}
\right\} =\beta ^{\prime }+\frac{1}{u};
\end{equation}

\begin{equation}
\,\,\,\,\,\,\,\,\,\,\,\left\{
\begin{array}{c}
3 \\
0\,\,3
\end{array}
\right\} =\left\{
\begin{array}{c}
3 \\
3\,\,\,0
\end{array}
\right\} =\left\{
\begin{array}{c}
3 \\
1\,\,\,\,3
\end{array}
\right\} =\left\{
\begin{array}{c}
3 \\
3\,\,\,\,1
\end{array}
\right\} =-\beta ^{\prime }+\frac{1}{u}.  \label{8}
\end{equation}

The non zero Ricci tensor components are $R_{00},R_{01}\,$and$\,R_{11}.$

\begin{equation}
R_{00}=\left\{
\begin{array}{c}
\mu \\
0\,\,\mu
\end{array}
\right\} _{,\mu }-(\ln \sqrt{g})_{,00}+(\ln \sqrt{g})_{,\mu }\left\{
\begin{array}{c}
\mu \\
0\,\,\,\,\,0
\end{array}
\right\} -\left\{
\begin{array}{c}
\mu \\
0\,\,\,\,\,\nu
\end{array}
\right\} \left\{
\begin{array}{c}
\nu \\
0\,\,\,\,\,\,\mu \,\,
\end{array}
\right\} .\,\,\,\,\,
\end{equation}

For the given metric (\ref{7}) we have

\begin{equation}
(\ln \sqrt{\mid g\mid })=2\ln u+2\Omega
.\,\,\,\,\,\,\,\,\,\,\,\,\,\,\,\,\,\,\,\,\,\,\,\,\,\,\,\,\,\,\,\,\,\,\,\,\,%
\,\,\,\,\,\,\,\,\,\,\,\,\,\,\,\,\,\,\,\,\,\,\,\,\,\,\,\,\,\,\,\,\,\,\,\,\,\,%
\,\,\,\,\,\,\,\,\,\,\,\,\,\,\,\,\,\,\,\,\,\,\,\,\,\,\,\,\,\,\,\,\,\,\,\,\,\,%
\,\,\,\,\,\,\,\,\,\,\,\,\,\,\,\,\,\,\,\,\,\,\,\,\,\,\,\,\,\,\,\,\,\,\,\,\,\,%
\,\,\,\,
\end{equation}

\begin{equation}
\ln \sqrt{\mid g\mid })^{\prime }=2(\Omega ^{\prime }+\frac{1}{u}
\end{equation}

\begin{equation}
(\ln \sqrt{\mid g\mid })^{\prime \prime }=2(\Omega ^{\prime \prime }-\frac{1%
}{u^{2}})
\end{equation}

Now using the Christoffel symbols listed in equation (\ref{8}) we get.

\begin{eqnarray*}
R_{00} &=&\left\{
\begin{array}{c}
0 \\
0\,\,0
\end{array}
\right\} _{,0}+\left\{
\begin{array}{c}
1 \\
0\,\,0
\end{array}
\right\} _{,1}-(\ln \sqrt{g})_{,00}+(\ln \sqrt{g})_{,0}\left\{
\begin{array}{c}
0 \\
0\,\,\,\,\,0
\end{array}
\right\} \\
&&-\left\{
\begin{array}{c}
0 \\
0\,\,\,\,\,0
\end{array}
\right\} +(\ln \sqrt{g})_{,1}\,\left\{
\begin{array}{c}
1 \\
0\,\,\,\,\,\,\,0
\end{array}
\right\} \,-2\left\{
\begin{array}{c}
0 \\
0\,\,\,\,\,1\,
\end{array}
\right\} \left\{
\begin{array}{c}
1 \\
0\,\,\,\,\,\,0\,\,
\end{array}
\right\} \\
&&\,\,\,\,\,\,\,\,\,\,\,\,\,\,\,\,\,\,\,\,\,\,\,\,\,\,\,\,\,\,\,\,\,\,\,\,\,%
\,\,\,\,\,\,\,\,\,\,\,\,\,\,\,\,\,\,\,\,\,\,\,\,\,\,\,\,\,\,\,\,\,\,\,\,\,\,%
\,\,\,\,\,\,\,\,\,\,\,\,\,\,-\left\{
\begin{array}{c}
2 \\
0\,\,\,\,\,\,2\,\,
\end{array}
\right\} ^{2}-\left\{
\begin{array}{c}
3 \\
0\,\,\,\,\,\,3\,\,
\end{array}
\right\} ^{2}\,.\,\,\,
\end{eqnarray*}
\begin{equation}
R_{00}=4\Omega ^{\prime }/u-2\beta ^{\prime
2}.\,\,\,\,\,\,\,\,\,\,\,\,\,\,\,\,\,\,\,\,\,\,\,\,\,\,\,\,\,\,\,\,\,\,\,\,%
\,\,\,\,\,\,\,\,\,\,\,\,\,\,\,\,\,\,\,\,\,\,\,\,\,\,\,\,\,\,\,\,\,\,\,\,\,\,%
\,\,\,\,\,\,\,\,\,\,\,\,\,\,\,\,\,\,\,\,\,\,\,\,\,\,\,\,\,\,\,\,\,\,\,\,\,\,%
\,\,\,\,\,\,\,\,\,\,\,\,\,\,\,\,\,
\end{equation}

Here the vacuum Einstein equation reduces to

\begin{equation}
4\frac{\Omega ^{\prime }}{u}-2\beta ^{\prime
2}=0.\,\,\,\,\,\,\,\,\,\,\,\,\,\,\,\,\,\,\,\,\,\,\,\,\,\,\,\,\,\,\,\,\,\,\,%
\,\,\,\,\,\,\,\,\,\,\,\,\,\,\,\,\,\,\,\,\,\,\,\,\,\,\,\,\,\,\,\,\,\,\,\,\,\,%
\,\,\,\,\,\,\,\,\,\,\,\,\,\,\,\,\,\,\,\,\,\,\,\,\,\,\,\,\,\,\,\,\,\,\,\,\,\,%
\,\,\,\,\,\,\,\,\,\,\,\,\,\,\,\,\,\,\,\,\,\,\,\,\,\,\,\,\,\,\,\,\,\,\,\,\,\,%
\,\,\,\,\,\,\,  \label{9}
\end{equation}

It is an exact plane gravitational wave solution.

\begin{equation}
R_{01}=-4\frac{\Omega ^{\prime }}{u}+2\beta ^{\prime
2}=0.\,\,\,\,\,\,\,\,\,\,\,\,\,\,\,\,\,\,\,\,\,\,\,\,\,\,\,\,\,\,\,\,\,\,\,%
\,\,\,\,\,\,\,\,\,\,\,\,\,\,\,\,\,\,\,\,\,\,\,\,\,\,\,\,\,\,\,\,\,\,\,\,\,\,%
\,\,\,\,\,\,\,\,\,\,\,\,\,\,\,\,\,\,\,\,\,\,\,\,\,\,\,\,\,\,\,\,\,\,\,\,\,\,%
\,\,\,\,\,\,\,\,\,\,\,\,\,\,\,\,\,\,  \label{9a}
\end{equation}

\begin{equation}
R_{11}=4\frac{\Omega ^{\prime }}{u}-2\beta ^{\prime
2}=0.\,\,\,\,\,\,\,\,\,\,\,\,\,\,\,\,\,\,\,\,\,\,\,\,\,\,\,\,\,\,\,\,\,\,\,%
\,\,\,\,\,\,\,\,\,\,\,\,\,\,\,\,\,\,\,\,\,\,\,\,\,\,\,\,\,\,\,\,\,\,\,\,\,\,%
\,\,\,\,\,\,\,\,\,\,\,\,\,\,\,\,\,\,\,\,\,\,\,\,\,\,\,\,\,\,\,\,\,\,\,\,\,\,%
\,\,\,\,\,\,\,\,\,\,\,\,\,\,\,\,\,\,\,\,\,  \label{9b}
\end{equation}
Similarly equations (\ref{9a})\thinspace \thinspace and (\ref{9b}) reduces
to same value as equation (\ref{9}), where other components of Ricci tensor
are zero. Hence the only non-trivial exact solution of the plane
gravitational wave is,

\begin{equation}
\Omega ^{\prime }(u)=\frac{1}{2}u\beta ^{\prime 2}.
\end{equation}

\section{ Interaction of a particle with plane gravitational waves}

Here it will be shown that gravitational waves carry energy following the
method of Weber and Wheeler [3], considering the plane gravitational wave (%
\ref{7}) , which is discussed in the previous section.

As already mentioned, here we will show that plane gravitational waves carry
energy. It can be shown by analyzing the motion of a particle which is
initially at rest and interacts with the plane gravitational wave. Write the
geodesic equation$.$

\begin{equation}
\stackrel{\cdot \cdot }{x}^{\mu }+\left\{
\begin{array}{c}
\mu \\
\nu \,\,\,\,\,\rho
\end{array}
\right\} \stackrel{\cdot }{x}^{\nu }\stackrel{\cdot }{x}^{\rho }=0.
\label{11}
\end{equation}
Using the Christoffel symbols, listed in (\ref{8}), and the usual summation
convention over the repeated indices, equation (\ref{11}) becomes:

\[
\frac{d^{2}x}{ds^{2}}+\Omega ^{\prime }(\frac{dt}{ds})^{2}+2\Omega ^{\prime }%
\frac{dt}{ds}\frac{dx}{ds}-\Omega ^{\prime }(\frac{dx}{ds})^{2}+u(u\beta
^{\prime }+1)e^{2(\beta -\Omega )}(\frac{ds}{dy})^{2}
\]
\begin{equation}
\,\,\,\,\,\,\,\,\,\,\,\,\,\,\,\,\,\,\,\,\,\,\,\,\,\,\,\,\,\,\,\,\,\,\,\,\,\,%
\,\,\,\,\,\,\,\,\,\,\,\,\,\,\,\,\,\,\,\,\,\,\,\,\,\,\,\,\,\,\,\,\,\,\,\,\,\,%
\,\,\,\,\,\,\,\,\,\,\,\,\,\,\,\,\,\,\,\,\,\,\,\,\,\,\,\,\,\,\,+u(-u\beta
^{\prime }+1)e^{-2(\beta +\Omega )}(\frac{dz}{ds})^{2}=0.  \label{12}
\end{equation}
Here in the above equation (\ref{12}) $\frac{dt}{ds}$ is unknown. It can
easily be found as follows. Dividing both sides of (\ref{7}) by $ds^{2}$ and
simplifying, we get

\begin{equation}
\frac{dt}{ds}=[(1+(-2\Omega +2\Omega ^{2}-\frac{4}{3}\Omega ^{3}+...+(\frac{%
dx}{ds})^{2}]^{1/2}.\,\,\,\,\,\,\,\,\,\,\,\,\,\,\,\,\,\,\,\,\,\,\,\,\,\,\,\,%
\,\,\,\,\,\,\,\,\,\,\,\,\,\,\,\,\,
\end{equation}
Using the binomial expansion, we get the following after further
simplification.

\begin{equation}
\frac{dt}{ds}=1-\Omega +\Omega ^{2}-\frac{2}{3}\Omega ^{3}+...+(\frac{dx}{ds}%
)^{2}+...\,\,\,\,\,\,\,\,\,\,\,\,\,\,\,\,\,\,\,\,\,\,\,\,\,\,\,\,\,\,\,\,\,%
\,\,\,\,\,\,\,\,\,\,\,\,\,\,\,\,\,\,\,\,\,\,\,  \label{13}
\end{equation}
Consider $\Omega $ as a first order quantity and $\beta $ a second order
quantity. Let
\begin{equation}
x=x_{(0)}+x_{(1)}+x_{(2)}+...\,\,\,\,\,\,\,\,\,\,\,\,\,\,\,\,\,\,\,\,\,\,\,%
\,\,\,\,\,\,\,\,\,\,\,\,\,\,\,\,\,\,\,\,\,\,\,\,\,\,\,\,\,\,\,\,\,\,\,\,\,\,%
\,\,\,\,\,\,\,\,\,\,\,\,\,\,\,\,\,\,\,\,\,\,\,\,\,\,\,\,\,\,\,\,\,\,\,\,\,\,%
\,\,\,\,\,\,\,\,\,\,\,\,\,\,\,\,\,\,\,
\end{equation}
Using these considerations and equation (\ref{13}), equation(\ref{12}) gives
the following approximation equations.

\begin{equation}
\frac{d^{2}x_{(0)}}{ds^{2}}+2\Omega ^{\prime }\frac{dt}{ds}\frac{dx_{(0)}}{ds%
}=0.\,\,\,\,\,\,\,\,\,\,\,\,\,\,\,\,\,\,\,\,\,\,\,\,\,\,\,\,\,\,\,\,\,\,\,\,%
\,\,\,\,\,\,\,\,\,\,\,\,\,\,\,\,\,\,\,\,\,\,\,\,\,\,\,\,\,\,\,\,\,\,\,\,\,\,%
\,\,\,\,\,  \label{14}
\end{equation}

\begin{equation}
\frac{d^{2}x_{(1)}}{ds^{2}}+2\Omega ^{\prime }\frac{dt}{ds}\frac{dx_{(1)}}{ds%
}-2\Omega ^{\prime }\frac{dx_{(0)}}{ds}\frac{dx_{(1)}}{ds}%
=0.\,\,\,\,\,\,\,\,\,\,\,\,\,\,\,\,\,\,\,\,\,  \label{15}
\end{equation}

\begin{equation}
\frac{d^{2}x_{(2)}}{ds^{2}}+2\Omega ^{\prime }(\frac{dt}{ds}-\frac{dx_{(0)}}{%
ds})\frac{dx_{(2)}}{ds}-2\Omega ^{\prime }(\frac{dx_{(1)}}{ds})^{2}=0.
\label{16}
\end{equation}

Equations (\ref{14}), (\ref{15}) and (\ref{16}) are zero order, first order
and second order approximation equations respectively. With the previously
developed tools we can integrate the approximation equations. Here it should
also be noted that as $\Omega $ is a first order quantity, its derivative
will also be a first order quantity. The result is:

$i)$ The zero order approximation

\begin{equation}
\frac{dx_{(0)}}{ds}=-2\Omega ^{\prime }(1-\Omega
)x_{(0)}+c_{1},\,\,\,\,\,\,\,\,\,\,\,\,\,\,\,\,\,\,\,\,\,\,\,\,\,\,\,\,\,\,%
\,\,\,\,\,\,\,\,\,\,\,\,\,\,  \label{17}
\end{equation}

$ii)$ The first order approximation

\begin{equation}
\frac{dx_{(1)}}{ds}=-2c_{2}\Omega ^{\prime
}x_{(1)}+c_{3};\,\,\,\,\,\,\,\,\,\,\,\,\,\,\,\,\,\,\,\,\,\,\,\,\,\,\,\,\,\,%
\,\,\,\,\,\,\,\,\,\,\,\,\,\,\,\,\,\,\,\,\,\,\,\,\,\,  \label{18}
\end{equation}

$iii)$ The second order approximation

\begin{equation}
\frac{dx_{(2)}}{ds}=-2c_{4}\Omega ^{\prime }x_{(2)}+2c_{5}\Omega ^{\prime
}s+c_{6}.\,\,\,\,\,\,\,\,\,\,\,\,\,\,\,\,\,\,\,\,\,\,\,\,\,\,\,\,  \label{19}
\end{equation}
Thus equations (\ref{17}), (\ref{18}) and (\ref{19}) guarantee that a
non-zero momentum is imparted to the particle after interacting with the
plane gravitational wave. Hence gravitational waves carry energy.

\section{ Sources of gravitational waves}

\subsection{ Spinning rod}

A rod spinning about an axis perpendicular to its length is one of the first
sources of gravitational radiation ever to have been considered [13].

Consider a steel beam of radius $r=1meter$ length $l=20\,meter,$ density $%
\rho =7.8gm/cm^{3}$, mass $M=4.9\times 10^{8}gm$ and tensile strength $%
t=40,000lbft/in^{2}$. Let the beam rotates about its middle, so it rotates
end over end with an angular velocity $\omega \,$\thinspace limited by the
balance centrifugal force and tensile strength.

\begin{equation}
\omega =(\frac{8t}{\rho l^{2}})^{1/2}=28radians/\sec
\textrm{\thinspace
\thinspace }\,\,\,\,\,\,\,\,\,\,\,\,\,\,\,\,\,\,\,\,\,\,\,\,\,\,\,\,\,\,\,\,%
\,\,\,\,\,\,\,\,\,\,\,\,\,\,\,\,\,\,\,\,\,\,\,\,\,
\end{equation}
The internal power flow is

\begin{eqnarray}
L_{internal} &=&(\frac{1}{2}I\omega ^{2})\omega =\frac{1}{28}Ml^{2}\omega
^{3}  \nonumber \\
&\approx &2\times 10^{8}erg/\sec \approx
10^{-41}L_{0}\,\,joule\,\,\,\,\,\,\,\,\,\,\,\,\,\,\,\,
\end{eqnarray}
where $L_{0}=c^{5}/G.\,$The order of magnitude of the power radiated is only
$L_{GW}\sim (10^{-41})^{2}L_{0}\sim 10^{-23}erg/\sec .$ Evidently the
construction of a laboratory generator of gravitational wave is
unattractive, as this is a very small quantity which cannot be detected
without new engineering or new ideas. There are however a great variety of
astrophysical sources of gravitational waves. We list some of them and then
discuss them lightly without going into actual calculation.

\subsection{ Astrophysical sources of gravitational waves}

Astrophysical sources of gravitational waves are the following{ :}

\begin{enumerate}
\item
\begin{enumerate}
\item  Pulsar

\item  Double star system;

\item  Gravitational collapse of a few solar mass star;

\item  Formation of a large black hole;.\thinspace \thinspace
\end{enumerate}
\end{enumerate}

Now let me explain them.

$\left( {\bf a}\right) ${\bf \ }{\Large Gravitational radiation from a pulsar%
}

Consider a highly dynamic astrophysical system. In particular take it to be
a wildly rotating pulsar. If its mass is $M$ and its size is $R$ then by
virial theorem its kinetic energy is $\sim M^{2}/R^{2}$ . The characteristic
time scale for mass to move from one side of the system to the other is

\begin{equation}
T\sim R/(mean\,velocity)\sim
R/(M/R)^{1/2}=(R^{3}/M)^{1/2}\,\,\,\,\,\,\,\,\,\,\,\,\,\,\,\,\,\,\,\,\,\,\,%
\,\,\,\,\,\,\,\,\,\,\,\,\,\,\,\,\,\,\,\,\,\,\,\,\,\,
\end{equation}
The internal power flow is

\begin{equation}
L_{int}\sim \frac{kinetic\,energy}{T}\sim (\frac{M^{2}}{R})(\frac{M^{2}}{%
R^{3}})^{1/2}\sim
(M/R)^{5/2}.\,\,\,\,\,\,\,\,\,\,\,\,\,\,\,\,\,\,\,\,\,\,\,\,\,\,\,\,\,\,\,\,%
\,\,\,\,\,\,\,\,
\end{equation}
The gravitational wave output is the square of this quantity or $L_{GW}\sim (%
\frac{M}{R})^{5}L_{0}.$ Clearly the maximum power output occurs when the
system is near its gravitational radius, and because nothing, not even
gravitational waves can escape from inside the gravitational radius. The
maximum value of the output is $\sim L_{0}=3.63\times 10^{59}erg/\sec $
regardless of the nature of the system.

$\left( {\bf b}\right) $ {\Large Double star system}

It has been estimated that at least one-fifth of all the stars are binary
systems. We will go into the details of how it happens so frequently because
that is a topic of astrophysics and hydrodynamics.

Consider two stars of masses $m_{1\,\textrm{ }}$and $m_{2}$
revolving in a circular orbit about their common centre of
gravity. For their circular frequency of revolution, we have the
standard formula:
\begin{equation}
\omega ^{2}=(m_{1+}m_{2)}/r^{3}\textrm{ (geometrical units of mass
and time)}
\end{equation}
The calculated rate of loss of energy by radiation is [13].

\begin{equation}
-\frac{dE}{dt}=(\frac{32}{5})(m_{1}m_{2}/(m_{1}+m_{2}))^{2}r^{4}\omega
^{6}\,\,\,\,\,\,\,\,\,\,\,\,\,\,\,\,\,\,\,\,\,\,\,\,\,\,\,\,\,\,\,\,\,\,\,\,%
\,\,\,\,\,\,\,\,\,\,\,\,\,
\end{equation}
The following are important types of double star systems.

$
\begin{array}{lll}
\textrm{Type name} & -(\frac{dE}{dt})_{grav} & \textrm{on Earth} \\
\eta \,\cos & 5.2\times 10^{10}erg/\sec & 1.4\times 10^{-29}erg/cm^{2}\sec
\\
\xi \,\beta oo & 3.6\times 10^{12}erg/\sec & 6.7\times 10^{-28}erg/cm^{2}\sec
\\
Wuma & 4.7\times 10^{29}erg/\sec & 3.2\times 10^{-13}erg/cm^{2}\sec
\end{array}
$

$\left( {\bf c}\right) $ \smallskip {\Large Gravitational collapse of a few
solar\thinspace \thinspace \thinspace \thinspace masses star}

Collapse to form neutron stars or black holes in the mass range $1$ to $%
10\,M_{\odot }$ $(M_{\odot }=2\times 10^{33}gm$ is a solar mass$)$ will
radiate waves in the frequency range 1 to 10 kHz with an amplitude that
depends on how much symmetry there is in the collapse. These collapses, at
least sometimes, result in supernova explosions. The rate at which supernova
occurs is relatively well known, but the fraction of collapse events that
produce strong enough gravitational waves is not well known. The
characteristic period of the waves is proportional to the light-travel time
around the collapsed object, the dominant frequency scale is $1/M$. For
sufficiently large $M$, the source will produce low frequency waves
detectable in space.

$\left( {\bf d}\right) ${\bf \ }{\Large Formation of a giant black holes}

Many astrophysicist believe that the most plausible explanation for quasars
and active galactic nuclei is that they contain massive ( $%
10^{6}-10^{9}M_{\odot })\,$black holes that accrete gas and stars to fuel
their activity. There is growing evidence that even so called normal
galaxies, like our own and Andromeda, contains black holes of modest size $%
(10^{4}-10^{6}M_{\odot })\,$in their nuclei. It is not clear how such holes
form, but if they form by the rapid collapse of a cluster of stars or of a
single supermassive star, then with a modest degree of non-symmetry in
collapse, they could produce amplitudes $h\approx 10^{-16}\,to\,\,10^{-18}$%
meters in the low frequency range observable from space. If a detector has
spectral noise density of $10^{-20}H_{z}^{-\frac{1}{2}}$then such events
could have signal to noise ratio ($\frac{S}{N})\,$of as much as 1000. This
strong signal would permit a detailed study of the event. If every galaxy
has one such black hole formed in this way , then there could be one event
per year in a galaxy. If no such events are seen, then either giant black
holes don't exist or they form much more gradually or with too much
spherical symmetry.

\chapter{REVIEW OF THE PSEUDO-NEWTONIAN FORMALISM AND ITS EXTENSION}

In the first section a review of the Pseudo-Newtonian $(\psi N)$ formalism
is given. Section 2 provides a review of the extension of the $\psi N$
formalism. In section 3 the extended $\psi N\,\,(e\psi N)$ formalism is used
to develop a formula for the momentum imparted to test particles in
arbitrary spacetime. Finally this formula is applied to plane and
cylindrical gravitational waves, both of which give very reasonable results.

\section{{ The $\Psi N$ formalism}}

The $\Psi N$ formalism [7] is based on the observation that, whereas the
gravitational force is not detectable in a freely falling frame (FFF), that
is so only at a point. It is detectable over a finite spatial extent as the
tidal force. It could be measured by an accelerometer, as shown in fig.3.1.

This accelerometer has a spring of length $``l"$ which connects two masses.
The spring ends in a needle which can move on the dial of the accelerometer
to give a measure of the tension in the spring. Thus an observer in the FFF
can observe the position of the spring by observing the moment of the needle
on the dial. In Newtonian gravitation theory, the needle will show a zero
position of the accelerometer in the absence of a central force. The force
exerted by the source pulls the mass near it more than the mass further
away. Thus the spring is stretched and the needle will move in the positive
direction.

If the spring is compressed the needle will show a negative deflection
otherwise it will show a positive deflection. This would occur if both discs
and the source consisted of like electric charges. Hence the negative
deflection corresponds to a repulsive source and the positive deflection
corresponds to an attractive source. The strength of the source would be
shown by the extent that the needle moved.

Mathematically the tidal acceleration is given by

\begin{equation}
A^{\mu }=R_{\nu \rho \pi }^{\mu }t^{\nu }l^{\rho }t^{\pi
}\,\,\,\,\,\,\,\,\,\,\,\,\,\,\,\,\,\,\,\,\,\,\,\,\,\,\,\,\,\,\,\,\,\,\,\,\,%
\,\,\,\,\,\,\,\,\,\,\,\,\,\,\,\,\,\,\,\,\,\,\,\,\,(\,\mu ,\nu
,...=0,1,2,3)\,\,\,\,\,\,\,\,\,\,\,\,\,\,\,\,\,\,\,\,\,\,\,\,\,\,\,\,\,\,\,%
\,\,
\end{equation}
$\,\,\,\,$Here ${\bf R}\,$is the Riemann tensor, ${\bf t}$ a
timelike killing vector, ${\bf l}$ is the spacelike separation
vector representing accelerometer. Thus in geometrical terms tidal
force is given by

\begin{figure}
\centerline{\epsfig{file=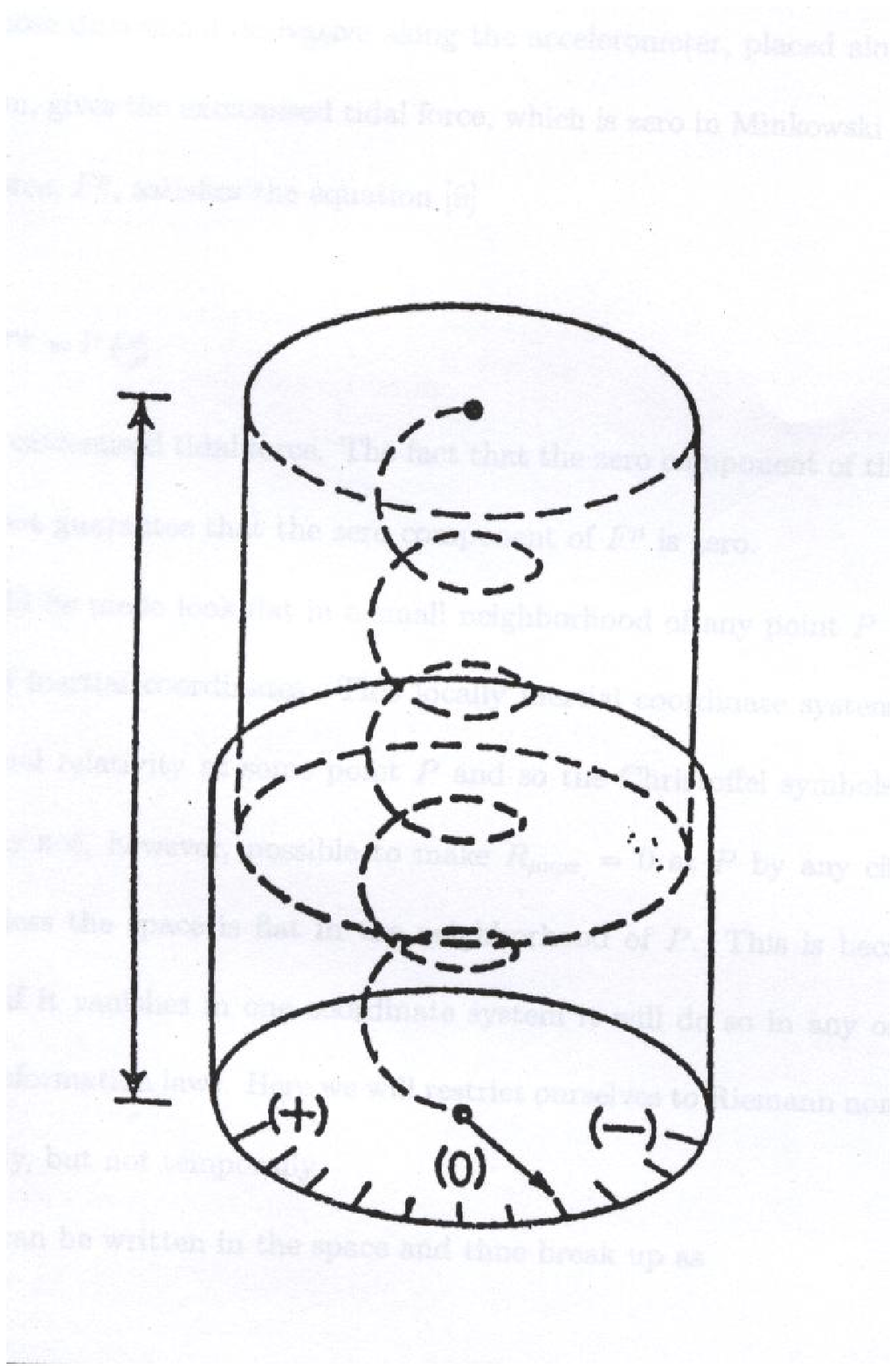,height=13cm,angle = 0}}
 \caption{An accelorometer to measure the
force due to a gravitating source. The force is measured by the
instrument in terms of the tension in the spring and shown by
motion of the needle on the dial.}
 \label{}
\end{figure}

\begin{equation}
F^{\mu }=-mR_{\nu \rho \pi }^{\mu }t^{\nu }l^{\rho }t^{\pi
}\,\,\,\,\,\,\,\,\,\,\,\,\,\,\,\,\,\,\,\,\,\,\,\,\,\,\,\,\,\,\,\,\,\,\,\,\,%
\,\,\,\,\,\,\,\,\,\,\,\,\,\,\,\,\,\,\,\,\,\,\,\,\,\,\,\,\,\,\,\,\,\,\,\,\,\,%
\,\,\,\,\,\,\,\,\,\,\,  \label{00.1}
\end{equation}
$\,\,\,\,\,\,\,\,\,\,\,\,\,\,\,\,\,\,\,\,\,\,\,\,\,$where $m\,$is
the mass of the test particle, $t^{\nu \textrm{ }}$the timelike
vector tangent to the particles path and $l^{\rho }$ the
separation vector, which provides the observation of the tidal
force. In the FFF $t^{\nu }=f\delta _{0}^{\nu
},\,\,\,\,f^{2}=\frac{1}{g_{00}}.$ Thus equation (\ref{00.1})
becomes:

\begin{equation}
F^{\mu }=-mf^{2}R_{0\rho 0}^{\mu }l^{\rho
}.\,\,\,\,\,\,\,\,\,\,\,\,\,\,\,\,\,\,\,\,\,\,\,\,\,\,\,\,\,\,\,\,\,\,\,\,\,%
\,\,\,\,\,\,\,\,\,\,\,\,\,\,\,\,\,\,\,\,\,\,\,\,\,\,\,\,\,\,\,\,\,\,\,\,\,\,%
\,\,\,\,\,  \label{00.2}
\end{equation}
Regarding this as an eigenvalue equation, we get the maximum tidal force
along the eigenvector of the matrix $R_{0\rho 0}^{\mu }$. Since $l^{\rho }$
is a purely spacelike vector in the free fall frame, the maximum tidal force
will be a purely space like vector. Thus we get
\begin{equation}
\stackrel{\_}{F}^{i}=\stackrel{\_}{l}^{j}\gamma
_{;j}^{i}\,\,\,\,\,\,\,\,\,\,\,\,\,\,\,\,\,\,\,\,\,\,\,\,\,\,\,\,\,\,\,%
\,(i,j=1,2,3)\,\,\,\,\,\,
\end{equation}
where `` ; '' stands for covariant derivative. Thus $\gamma ^{i}$
is the relativistic analogue of the Newtonian gravitational force.
Further this force is the gradient of a scalar quantity [7]
\begin{equation}
\gamma ^{i}=-\phi
_{,i}.\,\,\,\,\,\,\,\,\,\,\,\,\,\,\,\,\,\,\,\,\,\,\,\,\,\,\,\,\,\,\,\,\,\,\,%
\,\,\,\,\,\,\,\,\,\,\,\,\,\,\,\,\,\,\,\,\,\,\,\,\,\,\,\,\,\,\,\,\,\,\,\,\,\,%
\,\,\,\,\,
\end{equation}
Where an appropriate expression for $\phi $ is

\begin{equation}
\phi =\frac{1}{2}(g_{00}-1).\,\,\,\,\,\,\,\,\,\,\,\,\,\,\,\,\,\,\,\,\,\,\,\,%
\,\,\,\,\,\,\,\,\,\,\,\,\,\,\,\,\,\,\,\,\,\,\,\,\,\,\,\,\,\,\,\,\,\,\,\,\,\,%
\,\,\,\,\,\,\,\,\,\,\,\,\,\,\,\,\,\,
\end{equation}
When expressed in this way a Lorentz factor has to be introduced by hand. $%
\gamma ^{i}\,$and$\,\phi \,$are called the $\psi N$ force and
potential respectively.

\section{ The $e\protect\psi N$ force}

The quantity whose directional derivative along the accelerometer, placed
along the principal direction, gives the extremised tidal force, which is
zero in Minkowski space. Thus the $e\psi N$ force, $F^{\mu }$, satisfies the
equation [9]

\begin{equation}
F^{*\mu }=l^{\nu }F_{;\nu }^{\mu
}\,\,\,\,\,\,\,\,\,\,\,\,\,\,\,\,\,\,\,\,\,\,\,\,\,\,\,\,\,\,\,\,\,\,\,\,\,%
\,\,\,\,\,\,\,\,\,\,\,\,\,\,\,\,\,\,\,\,\,\,\,\,\,\,\,\,\,\,\,\,\,\,\,\,\,\,%
\,\,\,\,\,\,\,\,\,\,\,\,\,\,\,\,\,\,\,\,\,\,\,\,\,\,\,\,\,\,\,\,
\label{00.3}
\end{equation}
Where $F^{*\mu }$ is the extremised tidal force. The fact that the zero
component of the left side is zero does not guarantee that the zero
component of $F^{\mu }$ is zero.

The space could be made look flat in a small neighborhood of any point $P$
in by a special choice of inertial coordinates. This locally inertial
coordinate system has the metric of special relativity at some point $P$ and
so the Christoffel symbols $\Gamma _{\nu \rho }^{\mu }$ are zero at $P.$ It
is not, however, possible to make $R_{\mu \nu \rho \pi }=0$ at $P$ by any
choice of coordinates, unless the space is flat in the neighborhood of $P$.
This is because $R_{\mu \nu \rho \pi }$ is a tensor; if it vanishes in one
coordinate system it will do so in any other because of the transformation
laws. Here we will restrict ourselves to Riemann normal coordinates
spatially, but not temporally.

Equation (\ref{00.3}) can be written in the space and time break up as

\begin{equation}
l^{i}(F_{,i}^{0}+\Gamma
_{ij}^{0}F^{j})=0\,\,\,\,\,\,\,\,\,\,\,\,\,\,\,\,\,\,\,\,\,\,\,\,\,\,\,\,\,%
\,\,\,\,\,\,\,\,\,\,\,\,\,\,\,\,\,\,\,\,\,\,\,\,\,\,\,\,\,\,\,\,\,\,\,\,\,\,%
\,\,\,\,\,  \label{00.4}
\end{equation}

\begin{equation}
l^{j}(F_{,j}^{i}+\Gamma
_{0j}^{i}F^{0})=F^{*i}\,\,\,\,\,\,\,\,\,\,\,\,\,\,\,\,\,\,\,\,\,\,\,\,\,\,\,%
\,\,\,\,\,\,\,\,\,\,\,\,\,\,\,\,\,\,\,\,\,\,\,\,\,\,\,\,\,\,\,\,\,
\label{00.5}
\end{equation}
A simultaneous solution of the above equations can be obtained by using
Riemann normal coordinates for the spatial directions but not for the time
coordinates. Thus the $e\psi N$ force four vector is (see Appendix 2 for the
proof):

\begin{equation}
F^{0}=m[(\ln A)_{,0}-\Gamma _{00}^{0}+\Gamma _{0j}^{i}\Gamma
_{0i}^{j}/A]f^{2},\,\,\,\,\,\,\,\,\,\,\,\,\,\,\,\,  \label{00.5a}
\end{equation}

\begin{equation}
F^{i}=\Gamma
_{00}^{i}f^{2}.\,\,\,\,\,\,\,\,\,\,\,\,\,\,\,\,\,\,\,\,\,\,\,\,\,\,\,\,\,\,%
\,\,\,\,\,\,\,\,\,\,\,\,\,\,\,\,\,\,\,\,\,\,\,\,\,\,\,\,\,\,\,\,\,\,\,\,\,\,%
\,\,\,\,\,\,\,\,\,\,\,\,\,\,\,\,\,\,\,\,\,\,\,\,\,  \label{00.5b}
\end{equation}

where $A=(\ln \sqrt{-g})_{,0}\,\,\,\,\,\,\,\,g=\det (g_{ij})$ For these,
block diagonalized metrics

$\Gamma _{00}^{0}=\frac{1}{2}g^{00}g_{00,0}\,\,\,\,\,\,\,\,\,\,\,\,\,\,\,\,%
\,\,\,\,\,,\,\,\,\,\,\,\,\Gamma _{00}^{i}=-\frac{1}{2}g^{ij}g_{00,j}$

$\Gamma _{0j}^{i}=\frac{1}{2}g^{ij}g_{jk,0}$

Thus, writing the covariant form of the $e\psi N$ force

\begin{equation}
F_{0}=m[(\ln
Af)_{,0}-g_{,0}^{ij}g_{ij,0}/4A],\,\,\,\,\,\,\,\,\,\,\,\,\,\,\,\,\,\,\,\,\,%
\,\,\,\,\,\,\,\,\,\,\,\,\,  \label{00.5c}
\end{equation}

\begin{equation}
F_{i}=m(\ln
f)_{,i}.\,\,\,\,\,\,\,\,\,\,\,\,\,\,\,\,\,\,\,\,\,\,\,\,\,\,\,\,\,\,\,\,\,\,%
\,\,\,\,\,\,\,\,\,\,\,\,\,\,\,\,\,\,\,\,\,\,\,\,\,\,\,\,\,\,\,\,\,\,\,\,\,\,%
\,\,\,\,\,\,\,\,\,\,\,\,\,\,\,\,\,
\end{equation}
It is worth considering the significance of the zero component of the $e\psi
N$ force, which is its major difference from the $\psi N$ force. In special
relativistic terms, which are relevant for discussing forces in a Minkowski
space, the zero component of the four vector force corresponds to a proper
rate of change of energy of the test particle. Further we know that, in
general, an accelerated particle either radiates or absorbs energy according
as $\frac{dE}{dt}$ is less than or greater than zero. Thus $F_{0}$, should
also correspond to energy-emission or absorption by the background
spacetime. This point will be discussed further in the next chapter.

\section{ General formula for the momentum imparted to test particles in
arbitrary spacetime}

There has been a debate whether gravitational waves really exist [3,4]. To
demonstrate the reality of gravitational waves Ehler and Kundt [4]
considered a sphere of test particles in the path of plane fronted
gravitational waves and showed that a constant momentum was imparted to the
test particles. This was latter extended by Weber and Wheeler [3] for
cylindrical gravitational waves. The $\psi N$ force and $e\psi N$ force are
based on an operational procedure embodying the same principle [6]. The
proper time integral of the force four vector will be the momentum four
vector. Here it will be verified that this procedure gives the Ehler-Kundt
result for plane fronted gravitational waves. When it is applied to
cylindrical gravitational waves it is found that the result so obtained is
physically reasonable and gives an exact expression for the momentum
imparted to test particles, corresponding to the approximation given by
Weber and Wheeler.

From Equations (\ref{00.5a}) and (\ref{00.5b}) the momentum four vector,$%
\,P_{\mu }$, is obtained as:

\begin{equation}
P_{\mu }=\int F_{\mu
}dt.\,\,\,\,\,\,\,\,\,\,\,\,\,\,\,\,\,\,\,\,\,\,\,\,\,\,\,\,\,\,\,\,\,\,\,\,%
\,\,\,\,\,\,\,\,\,\,\,\,\,\,\,\,\,\,\,\,\,\,\,\,\,\,\,\,\,\,\,\,\,\,\,\,\,\,%
\,\,\,\,\,\,\,\,\,\,\,\,\,\,  \label{00.8}
\end{equation}
Now we apply this formula to plane fronted gravitational waves and
cylindrical gravitational waves.

\subsection{ Plane-fronted gravitational waves}

The metric for Plane-fronted gravitational waves is

\begin{equation}
ds^{2}=dt^{2}-dx^{2}-L^{2}(t,x)\{\exp (2\beta (t,x))dy^{2}+\exp (-2\beta
(t,x))dz^{2}\}.
\end{equation}
Where $L\,$and $\beta $ are arbitrary functions.

The metric tensor is

\begin{equation}
g_{ab}=\left(
\begin{array}{llll}
1 & 0 & 0 & 0 \\
0 & -1 & 0 & 0 \\
0 & 0 & -L^{2}e^{2\beta } & 0 \\
0 & 0 & 0 & -L^{2}e^{-2\beta }
\end{array}
\right) .
\end{equation}
Its inverse is

\begin{equation}
g^{ab}=\left(
\begin{array}{llll}
1 & 0 & 0 & 0 \\
0 & -1 & 0 & 0 \\
0 & 0 & -L^{-2}e^{-2\beta } & 0 \\
0 & 0 & 0 & -L^{-2}e^{2\beta }
\end{array}
\right) .
\end{equation}
To find the momentum four vector we need the force four vector. We calculate
the force four vector. Here in this special case:

\begin{equation}
g=\det (g_{ij})=-L^{4}(t,x),\,\,\,f=\frac{1}{\sqrt{g_{00}}}=1\,\,
\end{equation}

\begin{equation}
\ln \sqrt{-g}=2\ln
L\,\,\,\,\,\,\,\,\,\,\,\,\,\,\,\,\,\,\,\,\,\,\,\,\,\,\,\,\,\,\,\,\,\,\,\,\,%
\,\,\,\,\,\,\,\,\,\,\,\,\,\,\,\,\,\,\,\,\,\,\,\,\,\,\,\,\,\,\,\,\,\,\,\,\,\,%
\,\,\,\,\,\,\,\,
\end{equation}
which implies that

\begin{equation}
Af=(\ln \sqrt{-g}),_{0}=2\frac{\stackrel{.}{L}}{L}.\,\,\,\,\,\,\,\,\,\,\,\,%
\,\,\,\,\,\,\,\,\,\,\,\,\,\,\,\,\,\,\,\,\,\,\,\,\,\,\,\,\,\,\,\,\,\,\,\,\,\,%
\,\,\,\,\,\,\,\,\,\,\,\,  \label{00.9}
\end{equation}

\begin{equation}
(\ln (Af))_{,0}=\frac{\stackrel{..}{L}}{L}-\frac{\stackrel{.}{L}}{L}%
\,\,\,\,\,\,\,\,\,\,\,\,\,\,\,\,\,\,\,\,\,\,\,\,\,\,\,\,\,\,\,\,\,\,\,\,\,\,%
\,\,\,\,\,\,\,\,\,\,\,\,\,\,\,\,\,\,\,\,\,\,\,\,\,\,\,\,\,\,\,\,\,
\label{00.10}
\end{equation}

\begin{equation}
g_{,0}^{ij}g_{ij,o}=-8(\stackrel{.}{\beta }^{2}+\stackrel{.}{L}%
^{2}).\,\,\,\,\,\,\,\,\,\,\,\,\,\,\,\,\,\,\,\,\,\,\,\,\,\,\,\,\,\,\,\,\,\,\,%
\,\,\,\,\,\,\,\,\,\,\,\,\,\,\,\,\,\,\,  \label{00.11}
\end{equation}
So from equations (\ref{00.9}), (\ref{00.10}), (\ref{00.11}) and (\ref{00.5a}%
) we get

\begin{equation}
F_{0}=m(\stackrel{..}{L}+\stackrel{.}{\beta }^{2}L)/L.\,\,\,\,\,\,\,\,\,\,\,%
\,\,\,\,\,\,\,\,\,\,\,\,\,\,\,\,\,\,\,\,\,\,\,\,\,\,\,\,\,\,\,\,\,\,\,\,\,\,%
\,\,\,\,\,\,\,\,\,\,\,\,\,\,\,\,\,
\end{equation}
From the vacuum Einstein equations we have (already discussed in chapter 2)

\begin{equation}
\stackrel{..}{L}+\stackrel{.}{\beta }^{2}L=0,
\end{equation}
.which implies that $F_{0}=0$ that is the zero component of the force
four-vector is zero. Also $F_{i}=0$ because ($\ln \sqrt{g_{00}})_{,i}=0.$

Thus the momentum four vector becomes $P_{\mu }=${\it constant}$.$ Hence
there is a constant energy and momentum imparted to the test particles. The
constant here determines the strength of the wave. This exactly coincides
with the Ehler-Kundt method in that they demonstrate that the test particles
acquires a constant momentum and hence a constant energy from a plane
gravitational wave.

\subsection{ Cylindrical gravitational waves}

Consider the cylindrically symmetric metric (\ref{2.1}). We first calculate
the force four vector. Here in this special case:

\begin{equation}
g=\det (g_{ij})=-e^{4(\gamma -\psi
)}.\,\,\,\,\,\,\,\,\,\,\,\,\,\,\,\,\,\,\,\,\,\,\,\,\,\,\,\,\,\,\,\,\,\,\,\,%
\,\,\,\,\,\,\,\,\,\,\,\,\,\,\,\,\,\,\,\,\,\,\,\,\,\,\,\,\,\,\,\,\,\,\,\,\,\,%
\,\,  \label{00.12}
\end{equation}

\begin{equation}
\ln \sqrt{-g}=2(\gamma -\psi
)\,.\,\,\,\,\,\,\,\,\,\,\,\,\,\,\,\,\,\,\,\,\,\,\,\,\,\,\,\,\,\,\,\,\,\,\,\,%
\,\,\,\,\,\,\,\,\,\,\,\,\,\,\,\,\,\,\,\,\,\,\,\,\,\,\,\,\,\,\,\,\,\,\,\,\,\,%
\,  \label{00.13}
\end{equation}

\begin{equation}
A=2(\gamma ^{.}-\psi ^{.}),\,\,\,\,\,\,\,\,\,f=e^{\psi -\gamma
}.\,\,\,\,\,\,\,\,\,\,\,\,\,\,\,\,\,\,\,\,\,\,\,\,\,\,\,\,\,\,\,\,\,\,\,\,\,%
\,\,\,\,\,\,\,\,\,\,\,\,\,\,\,\,\,\,\,\,\,\,\,\,\,  \label{00.14}
\end{equation}

\begin{equation}
(\ln (Af))_{,0}=\frac{(\gamma ^{..}-\psi ^{..})\,\,\,\,\,}{(\gamma ^{.}-\psi
^{.})\,\,\,\,\,}-(\gamma ^{.}-\psi
^{.}).\,\,\,\,\,\,\,\,\,\,\,\,\,\,\,\,\,\,\,\,\,\,\,\,\,\,\,\,\,\,\,\,\,\,\,%
\,\,\,\,\,  \label{00.15}
\end{equation}

\begin{equation}
g_{,0}^{ij}g_{ij,0}=-4(\gamma ^{.}-\psi ^{.})^{2}-8\psi
^{.2}.\,\,\,\,\,\,\,\,\,\,\,\,\,\,\,\,\,\,\,\,\,\,\,\,\,\,\,\,\,\,\,\,\,\,\,%
\,\,\,\,\,\,\,\,\,\,\,\,\,\,\,\,\,\,\,\,\,\,\,  \label{00.16}
\end{equation}
where a dot denotes differentiation with respect to $t$.

The vacuum Einstein field equation gives [3].

\begin{equation}
\psi (t,\rho )=AJ_{0}(\omega \rho )\cos (\omega t)+BN_{0}(\omega \rho )\sin
(\omega t),  \label{00.17}
\end{equation}
where $\,J_{0}(\omega \rho )\,$\thinspace and$\,\,N_{0}(\omega \rho )\,$ are
the zero order Basel and Neuman functions respectively. $A$ and $B$ are
arbitrary constants corresponding to the strength of gravitational wave.

\begin{eqnarray}
\gamma (t,\rho ) &=&\frac{1}{2}\omega \rho \{(A^{2}J_{0}(\omega \rho
)J_{0}^{\prime }(\omega \rho )-B^{2}N_{0}(x)N_{0}^{\prime }(x))\cos (2\omega
t)  \nonumber \\
&&-AB[(J_{0}(\omega \rho )N_{0}^{\prime }(\omega \rho )+J_{0}^{\prime
}(x)N_{0}(x))\sin (2\omega t)-  \nonumber \\
&&-2(J_{0}(\omega \rho )N_{0}^{\prime }(\omega \rho )-J_{0}^{\prime
}(x)N_{0}(x))\omega t]\}  \label{00.19}
\end{eqnarray}
Where prime now refers to differentiation with respect to $\omega \rho
,\,\omega $ being the angular frequency.

Using equations (\ref{00.15}), (\ref{00.16}), (\ref{00.17}) and (\ref{00.19}%
) we get the zero component of the force four vector.

\begin{eqnarray}
F_{0} &=&-m\omega \{[AJ_{0}\cos (\omega t)+BN_{0}\sin (\omega t)]-2\rho
\omega [(A^{2}J_{0}J_{0}^{\prime }-B^{2}N_{0}N_{0}^{\prime })\cos (2\omega t)
\nonumber \\
&&-AB(J_{0}N_{0}^{\prime }+N_{0}J_{0}^{\prime })\sin (2\omega t)]+\frac{%
2[AJ_{0}\sin (\omega t)-BN_{0}\cos (\omega t)]^{2}}{AJ_{0}\sin (\omega t)}
\nonumber \\
&&-BN_{0}\cos (\omega t)-2\omega \rho [AJ_{0}\sin (\omega t)-BN_{0}\cos
(\omega t)]  \nonumber \\
&&[AJ_{0}^{\prime }\cos (\omega t)+BN_{0}^{\prime }\sin (\omega t)]\}.
\label{00.20}
\end{eqnarray}
As $\gamma $ and $\psi $ are functions of $t$ and $\rho $ only so $F_{2}\,$%
and\thinspace $F_{3}$ are zero.

\begin{eqnarray}
F_{1} &=&m(\ln f)_{,1}=-m\omega \{[AJ_{0}^{\prime }\cos (\omega
t)-BN_{0}^{\prime }\sin (\omega t)-\frac{1}{2}[(A^{2}J_{0}J_{0}^{\prime
}-B^{2}N_{0}N_{0}^{\prime })  \nonumber \\
&&+\omega \rho (A^{2}J_{0}J_{0}^{\prime }-B^{2}N_{0}N_{0}^{\prime })^{\prime
}]\cos (2\omega t)-\frac{1}{2}AB[2(J_{0}N_{0}^{\prime }+J_{0}^{\prime
}N_{0})^{\prime }]\sin (2\omega t)  \nonumber \\
&&-\frac{1}{2}AB[4(J_{0}N_{0}^{\prime }-J_{0}^{\prime }N_{0})+2\omega \rho
(J_{0}N_{0}^{\prime }-J_{0}^{\prime }N_{0})^{\prime }]\omega t\}.
\end{eqnarray}
The corresponding $P_{0}$ and $P_{1}$ are

\newpage

\[
P_{0}=-m[\ln \mid AJ_{0}\sin (\omega t)-BN_{0}\cos (\omega t)\mid +\ln \mid
1-2\omega \rho [AJ_{0}^{\prime }\cos (\omega t)
\]
\[
+BN_{0}^{\prime }\sin (\omega t)]\,\mid \left( 1+\frac{A^{2}J_{0}J_{0}^{%
\prime }+B^{2}N_{0}N_{0}^{\prime }}{\omega \rho (AJ_{0}^{\prime
2}+BN_{0}^{\prime 2})}\right)
\,\,\,\,\,\,\,\,\,\,\,\,\,\,\,\,\,\,\,\,\,\,\,\,\,\,\,\,\,\,\,\,\,\,\,\,\,\,%
\,\,\,\,\,\,\,\,
\]
\[
-\frac{AB(AJ_{0}N_{0}N_{0}^{\prime }+N_{0}J_{0}^{\prime }-A\omega \rho
J_{0}J_{0}^{\prime }N_{0}^{\prime }-N_{0}^{\prime }J_{0})}{\omega \rho
(A^{2}J_{0}^{\prime 2}+B^{2}N_{0}^{\prime 2})\sqrt{1-4\omega ^{2}\rho
^{2}(A^{2}J_{0}^{\prime 2}+B^{2}N_{0}^{\prime 2})}}
\]
$\,\,\,\,\,\,\,\,\,$%
\begin{eqnarray}
\,\,\,\,\,\,\,\,\,\,\,\,\,\,\,\,\,\,\,\,\,\,\,\,\,\,\,\,\,\,\,\,\,\,\,\,\,%
\tan ^{-1} &\mid &\frac{(1+2A\omega \rho J_{0}^{\prime })\tan (\frac{1}{2}%
\omega t)-2B\omega \rho N_{0}^{\prime }}{\sqrt{1-4\omega ^{2}\rho
^{2}(A^{2}J_{0}^{\prime 2}+B^{2}N_{0}^{\prime 2})}}\mid +\frac{%
AB(N_{0}J_{0}^{\prime }-J_{0}N_{0}^{\prime }}{\rho (A^{2}J_{0}^{\prime
2}+B^{2}N_{0}^{\prime 2})}t  \nonumber \\
&&\,\,\,\,\,\,\,\,\,\,\,\,\,\,\,\,\,\,\,\,\,\,\,\,\,\,\,\,\,\,\,\,\,\,\,\,\,%
\,\,\,\,\,\,\,\,\,\,\,\,\,\,\,\,\,\,\,\,\,\,\,\,\,\,\,\,\,\,\,\,\,\,\,\,\,\,%
\,\,\,\,\,\,\,\,\,\,\,\,\,\,\,\,\,\,\,\,\,\,\,\,\,\,\,\,\,\,\,\,\,\,+f_{1}(%
\omega \rho ).\,\,\,  \label{00.21a}
\end{eqnarray}
$\,\,\,\,\,$%
\begin{eqnarray}
P_{1} &=&-m\{[AJ_{0}^{\prime }\sin (\omega t)-BN_{0}^{\prime }\cos (\omega
t)]-\frac{1}{4}[(A^{2}J_{0}J_{0}^{\prime }-B^{2}N_{0}N_{0}^{\prime })
\nonumber \\
&&+\omega \rho (A^{2}J_{0}J_{0}^{\prime }-B^{2}N_{0}N_{0}^{\prime })^{\prime
}]\sin (2\omega t)-\frac{1}{2}AB[(J_{0}N_{0}^{\prime }+J_{0}^{\prime }N_{0})
\nonumber \\
&&\omega \rho (J_{0}N_{0}^{\prime }+J_{0}^{\prime }N_{0})^{\prime }]\cos
(2\omega t)-AB\omega ^{2}t^{2}[(J_{0}N_{0}^{\prime }+J_{0}^{\prime }N_{0})
\nonumber \\
&&-\omega \rho (J_{0}N_{0}^{\prime }-J_{0}^{\prime }N_{0})^{\prime
}]\}+f_{2}(\omega \rho ).  \label{00.22}
\end{eqnarray}
Where $f_{1}$ and $f_{2}$ are arbitrary constants of integration. Weber and
Wheeler exclude solution that contain irregular Bessel function, $%
N_{0}(\omega \rho )$, as not well defined at the origin. Taking the
Weber-Wheeler solutions equations (\ref{00.21a}) and (\ref{00.22}) reduces to

\begin{eqnarray}
P_{0} &=&-m[\ln \mid AJ_{0}\sin (\omega t)\mid +(1+AJ_{0}/\omega \rho
J_{0}^{\prime })\ln \mid 1-2\omega \rho AJ_{0}^{\prime }\cos (\omega t)\mid
\label{00.23} \\
&&\,\,\,\,\,\,\,\,\,\,\,\,\,\,\,\,\,\,\,\,\,\,\,\,\,\,\,\,\,\,\,\,\,\,\,\,\,%
\,\,\,\,\,\,\,\,\,\,\,\,\,\,\,\,\,\,\,\,\,\,\,\,\,\,\,\,\,\,\,\,\,\,\,\,\,\,%
\,\,\,\,\,\,\,\,\,\,\,\,\,\,\,\,\,\,\,\,\,\,\,\,\,\,\,\,\,\,\,\,\,\,\,\,\,\,%
\,\,\,\,\,\,\,\,\,\,\,\,\,\,\,\,\,\,+f_{1}(\omega \rho ).  \nonumber
\end{eqnarray}

\begin{equation}
P_{1}=-m\{AJ_{0}^{\prime }\sin (\omega t)-\frac{1}{4}A^{2}J_{0}J_{0}^{\prime
}+\rho \omega (J_{0}J_{0}^{\prime })^{\prime }\sin (2\omega t)+f_{2}(\omega
\rho )\}.  \label{00.24}
\end{equation}
We see that the quantity $P_{1}$ given by equation (\ref{00.24}) can be made
zero for the large and small $\rho $ limits by choosing $\,f_{2}$ equal to
zero. This is physically reasonable expression for the momentum imparted to
test particles by cylindrical gravitational waves. The quantity $P_{0}$
given by equation (\ref{00.23}) remains finite for small $\rho $ and can
also be made finite for large $\rho $ by choosing $f_{1}=-\ln (J_{0})$.
However, there is a singularity at $\omega t=n\pi $. This problem does not
arise in the general expression given by equation (\ref{00.21a}). However,
in that case there appears a term linear in time which creates
interpretational problems. Also $P_{0}$ and $P_{1}$ becomes singular at $%
\rho =0$ if $B\neq 0$.

\chapter{ SPIN IMPARTED TO TEST PARTICLES BY GRAVITATIONAL WAVES}

We can write the $e\psi N$ force four vector (discussed in section 3.2) as $%
\left[ 7\right] $

\begin{equation}
F_{0}=-U_{,0}\,\,\,\,\,\,\,,\,\,\,\,\,\,\,\,\,\,\,\,\,\,\,\,\,\,\,\,\,\,\,\,%
\,\,\,\,\,\,\,\,\,F_{i}=-V_{,i}.
\end{equation}
where
\begin{equation}
U=m\left[ \ln \left( \frac{Af}{B}\right) -\int \frac{g_{,0}^{ij}g_{ij,0}}{4A}%
dt\right] ,
\end{equation}
\begin{equation}
V=-m\ln
(f),\,\,\,\,\,\,\,\,\,\,\,\,\,\,\,\,\,\,\,\,\,\,\,\,\,\,\,\,\,\,\,\,\,\,\,\,%
\,\,\,\,\,\,\,\,\,\,\,\,\,\,\,\,\,\,\,\,\,\,\,\,\,\,\,\,\,\,\,\,\,\,\,\,\,\,
\end{equation}
where B is a constant with units of time inverse so as to make $\frac{A}{B}$
dimensionless. Here $V$ is the $e\psi N\,$ potential but there is no good
interpretation of $U$. Thus there is a problem of interpretation of $F_{0}$.
The $F_{i}$ was reinterpreted $\left[ 8\right] $ as the rate of change of
the momentum imparted to test particles by the gravitational field, i.e. $%
F_{i}=\frac{dP_{i}}{d\tau }$ (where $\tau $ is the proper time). In this
interpretation it would be natural to identify $F_{0}\,$as $\frac{dP_{0}}{%
d\tau }$. The problem now is to interpret $P_{0}$ since one would normally
take $P_{0}=E=\sqrt{m^{2}+P^{i}P^{j}g_{ij}}$ where $m$ is the mass of the
test particle . Hence $\int F_{0}dt$ cannot be this $P_{0}$. Sharif's
suggestion $[11]$ for the interpretation of $P_{0},$ that it gives the spin
angular momentum imparted to test rods, is given in the first section of
this chapter, but in the same section it turns out to be inconsistent. To
find an alternative check on its validity the geodesic $\left[ 4\right] $
analysis for the angular momentum imparted to test particles by
gravitational waves is undertaken in section 2. This formalism is applied to
various cases in section 3.

\section{ Spin angular momentum imparted by gravitational waves}

Sharif [11]  considers a test rod of length $\lambda $ in the path
of a gravitational wave whose preferred direction is given by
$l^{i}$ in the preferred reference frame. He argues that the rod
will acquire
maximum angular momentum from the wave if it lies in the plane given by $%
e_{\rho jki}l^{\rho },$ where $e_{\mu \nu \rho \pi }$ is a totally skew
symmetric fourth rank tensor. Thus the spin vector will be given by

\begin{equation}
S^{i}=\frac{1}{2}e^{ijk\nu }e_{jkl}l^{l}P_{\nu }.
\end{equation}
For $i=1,$%
\begin{eqnarray}
S^{1} &=&\frac{1}{2}e^{1jk\nu }e_{jkl}l^{l}P_{\nu }  \nonumber \\
&=&\frac{1}{2}[e^{1230}e_{23l}+e^{1320}e_{32l}]l^{l}P_{0}  \nonumber \\
&=&P_{0}l^{1}.
\end{eqnarray}
Similarly

\begin{equation}
S^{2}=P_{0}l^{2}\textrm{ and }S^{3}=P_{0}l^{3}.\,\,\,\,\,\,\,\,\,\,\,\,\,\,\,%
\,\,\,\,\,\,\,\,\,\,\,\,\,\,\,\,\,\,\,\,\,\,\,\,\,\,
\end{equation}
So in the preferred direction the spin vector would be proportional to $%
l^{i} $ such that

\begin{equation}
S^{i}=P_{0}l^{i}.\;\;\;\;\;\;\;\;\;\;\;\;\;\;\;\;\;\;\;\;\;\;\;\;\;\;\;\;\;%
\;\;\;
\end{equation}
The angular momentum imparted would be the magnitude of the spin vector .
Thus the maximum angular momentum imparted to a test rod when it lies in the
plane perpendicular to the preferred direction is:

\begin{equation}
S=P_{0}\lambda =m\lambda \int [(\ln (Af)_{,0}-g_{,0}^{ij}g_{ij,0}/4A]dt.
\label{4.1}
\end{equation}
Hence the physical significance of the zero component of the momentum four
vector would be that it provides an expression for the spin imparted to a
test rod in an arbitrary spacetime.

This formula was applied to plane and cylindrical gravitational waves to
give the following results.\newpage

\smallskip

{\bf 1. Plane gravitational waves.}

\thinspace \thinspace \thinspace \thinspace \thinspace \thinspace \thinspace
\thinspace Using metric (\ref{7}) in equation (\ref{4.1}) we get (for
detailed calculations see section 3.2)
\begin{equation}
P_{0}=\textrm{{\it constant.}}
\end{equation}

and thus the spin would also be constant.

{\bf 2. Cylindrical gravitational waves.}

\thinspace \thinspace \thinspace \thinspace \thinspace \thinspace \thinspace
\thinspace \thinspace Following the same procedure for the metric (2.1) we
get (for detailed calculations see section 3.2)

\begin{equation}
S=-m\lambda [(1+AJ_{0}/\omega \rho J_{0}^{\prime })\ln \mid
1-2\omega \rho AJ_{0}^{\prime }\cos (\omega t)\mid +\textrm{{\it
constant.}}
\end{equation}

Notice that there can be no spin angular momentum imparted to test particles
in a perfectly homogeneous and isotropic cosmological model [1]; its high
degree of symmetry $-$ in particular, spherical symmetry is incompatible
with spin being imparted to test particles. However when we use Sharif's
formula for cosmological models, it gives exactly this error.

We give examples which had already partly been constructed by M. Sharif [14].

{\bf 3.\thinspace \thinspace The Friedman model}:

Consider the Friedman model, which is isotropic and homogeneous,

\begin{equation}
ds^{2}=dt^{2}-a^{2}(t)[d\chi ^{2}+f_{k}^{2}(\chi )d\Omega
^{2}],\,\,\,\,\,\,\,\,\,\,\,\,\,\,\,\,\,\,  \label{4.2}
\end{equation}

where

$k=1\,\,\,\,\,\,\,\,\,\,\,\,\,\,\,\,\,\,\,\,\,\,\,\,\,\,\,\,\,\,\,\,\,f_{1}(%
\chi )=\sin (\chi
),\,\,\,\,\,\,\,\,\,\,\,\,\,\,\,\,\,\,\,\,\,\,\,\,\,\,\,\,\,$

$k=0\,\,\,\,\,\,\,\,\,\,\,\,\,\,\,\,\,\,\,\,\,\,\,\,\,\,\,\,\,\,\,\,\,f_{0}(%
\chi )=\chi ,$

$k=-1\,\,\,\,\,\,\,\,\,\,\,\,\,\,\,\,\,\,\,\,\,\,\,\,\,\,\,f_{-1}(\chi
)=\sinh (\chi ),$

$\chi $ being the hyperspherical angle and $a(t)$ the scale parameter given
by:

\begin{equation}
\left.
\begin{array}{c}
a(t)=a_{0}(1-\cos \eta )/2,\,\,\,\,t=a_{0}(\eta -\sin \eta
),\,\,\,\,\,\,\,\,\,\,\,\,\,0\leq \eta \leq 2\pi ,\,\,k=1 \\
\,\,\,\,\,\,\,\,\,=a_{0}^{1/3}t^{2/3}\,\,\,\,\,\,\,\,\,\,\,\,\,\,\,\,\,\,\,%
\,\,\,\,\,\,\,\,\,\,\,\,\,\,\,\,\,\,\,\,\,\,\,\,\,\,\,\,\,\,\,\,\,\,\,\,\,\,%
\,\,\,\,\,\,\,\,\,\,\,\,\,\,\,\,\,\,\,\,\,\,\,\,\,\,\,\,\,\,\,\,\,\,\,\,\,\,%
\,\,\,\,\,k=0; \\
\,\,\,\,\,\,\,\,\,\,\,\,\,\,\,\,\,=a_{0}(\cosh \eta
-1)/2,\,\,\,\,t=a_{0}(\sinh \eta -\eta )/2,\,\,0\leq \eta \leq \infty
,\,\,\,k=-1;
\end{array}
\right\} \,\,\,\,\,\,  \label{a(t)}
\end{equation}
Using the metric (\ref{4.2}) the $e\psi N$ force for the Friedmann models
will be
\begin{equation}
F_{0}=-m\frac{a^{..}}{a^{.}},
\end{equation}
and

\begin{equation}
P_{0}=\int F_{0}dt=-m\ln (a^{.}).  \label{F0}
\end{equation}
As $S=m\lambda \int F_{0}dt.$ We get the spin for Friedmann models
\begin{equation}
S=-m\lambda \ln (a^{.}).  \label{S}
\end{equation}
{\bf a. Closed Friedmann model}

Using the metric (\ref{4.2}) with $k=1$ in equation (\ref{F0}) and (\ref{S})
we get
\begin{equation}
P_{0}=m[\ln \sqrt{1-\cos \eta }-\frac{3}{8}\cos \eta +\frac{1}{16}\cos
^{2}\eta +\ln \sqrt{2}+\frac{5}{16}],\,\,\,\,\,\,\,  \label{4.3a}
\end{equation}
and thus
\begin{equation}
S=m\lambda [\ln \sqrt{1-\cos \eta }-\frac{3}{8}\cos \eta +\frac{1}{16}\cos
^{2}\eta +\ln \sqrt{2}+\frac{5}{16}].  \label{4.3}
\end{equation}

{\bf b. Flat Friedmann model.}

Again Using the metric (\ref{4.2}) with $k=0$ in equation (\ref{F0}) and (%
\ref{S}) we get$\,\,\,\,\,\,\,$
\begin{equation}
P_{0}=m\ln \left( \frac{\eta }{2}\right) .  \label{4.4a}
\end{equation}
And thus the spin for a flat Friedmann model is
\begin{equation}
S=m\lambda \ln \left( \frac{\eta }{2}\right)
\,.\,\,\,\,\,\,\,\,\,\,\,\,\,\,\,\,\,\,\,  \label{4.4}
\end{equation}
{\bf c. Open Friedmann model.}

Finally we obtain the spin for the open Friedmann model by the same
procedure as the following
\begin{equation}
P_{0}=-m\ln \mid \frac{\sinh \eta }{\cosh \eta -1}\mid .\,  \label{4.5**}
\end{equation}
\begin{equation}
S=-m\lambda \ln \mid \frac{\sinh \eta }{\cosh \eta -1}\mid .\,  \label{4.5*}
\end{equation}
$.$

Equations (\ref{4.3}), (\ref{4.4}) and (\ref{4.5*}) tells us that a non-zero
spin is imparted by flat, open, and closed Friedmann models. Since no spin
can be imparted thus Sharif's interpretation of $P_{0}$ cannot be correct.

{\bf 4. The Kasner model}

Consider the Kasner model for a homogeneous anisotropic universe (near the
cosmological singularity)
\begin{equation}
ds^{2}=dt^{2}-t^{2p_{1}}dx^{2}-t^{2p_{2}}dy^{2}-t^{2p_{3}}dz^{2}  \label{k}
\end{equation}
Here $p_{i}$ are numbers such that $%
p_{1}+p_{2}+p_{3}=p_{1}^{2}+p_{2}^{2}+p_{3}^{2}=1.$

The metric tensor is
\begin{equation}
g_{\mu \nu }=\left(
\begin{array}{cccc}
1 & 0 & 0 & 0 \\
0 & -t^{2p_{1}} & 0 & 0 \\
0 & 0 & -t^{2p_{2}} & 0 \\
0 & 0 & 0 & -t^{2p3}
\end{array}
\right)
\end{equation}
Its inverse is
\begin{equation}
g^{\mu \nu }=\left(
\begin{array}{cccc}
1 & 0 & 0 & 0 \\
0 & -t^{-2p_{1}} & 0 & 0 \\
0 & 0 & -t^{-2p_{2}} & 0 \\
0 & 0 & 0 & -t^{-2p3}
\end{array}
\right)
\end{equation}

To find the momentum four vector we need the force four vector. We therefore
calculate it. Now

\begin{equation}
g=\det (g_{ij})=-t^{2},\,\,\,f=\frac{1}{\sqrt{g_{00}}}=1,\,\,
\end{equation}
\begin{equation}
\ln \sqrt{-g}=\ln
t,\,\,\,\,\,\,\,\,\,\,\,\,\,\,\,\,\,\,\,\,\,\,\,\,\,\,\,\,\,\,\,\,\,\,\,\,\,%
\,\,\,\,\,\,\,\,\,\,\,\,\,\,\,\,\,\,\,\,\,\,\,\,\,\,\,
\end{equation}
which implies that
\begin{equation}
Af=(\ln \sqrt{-g})_{,0}=\frac{1}{t},\,\,\,\,\,\,\,\,\,\,\,\,\,\,\,\,\,\,\,\,%
\,\,\,\,\,\,\,\,\,\,\,\,\,\,\,
\end{equation}
\begin{equation}
(\ln (Af))_{,0}=-\frac{1}{t},\,\,\,\,\,\,\,\,\,\,\,\,\,\,\,\,\,\,\,\,\,\,\,%
\,\,\,\,\,\,\,\,\,\,\,\,\,\,\,\,\,\,\,\,\,\,\,\,\,\,\,\,  \label{kas1}
\end{equation}
\begin{equation}
g_{,0}^{ij}g_{ij,o}=\frac{-4}{t^{2}}.\,\,\,\,\,\,\,\,\,\,\,\,\,\,\,\,\,\,\,%
\,\,\,\,\,\,\,\,\,\,\,\,\,\,\,\,\,\,\,\,\,\,\,\,\,\,\,\,\,\,\,\,\,\,\,
\label{kas2}
\end{equation}
So from equations (\ref{kas1}) and (\ref{kas2}) we get
\begin{equation}
F_{0}=0\,\,\,\,\,\,\,\,\,\,\,\,\,\,\,\,\,\,\,\,\,\,\,\,\,\,\,\,\,\,\,\,\,\,%
\,\,\,\,\,\,\,\,\,\,\,\,\,\,\,\,\,\,\,\,\,\,\,\,\,\,\,\,\,\,\,\,\,\,\,\,\,\,%
\,\,\,\,\,\,\,\,\,\,\,\,\,\,\,\,\,\,\,\,\,\,\,\,\,\,\,\,\,\,\,\,\,\,\,\,
\end{equation}
Hence from equation (\ref{4.1}) we get
\begin{equation}
S=\textrm{{\it constant.}}
\end{equation}
By physical consideration we could set $S=0.$

{\bf 5. The De Sitter universe (usual coordinates)}

\thinspace \thinspace \thinspace \thinspace \thinspace \thinspace \thinspace
\thinspace \thinspace Consider the metric
\begin{equation}
ds^{2}=(1-r^{2}/D^{2})dt^{2}-(1-r^{2}/D^{2})^{-1}dr^{2}-r^{2}(d\theta
^{2}+\sin ^{2}\theta d\varphi ^{2}),  \label{des}
\end{equation}

The metric tensor is
\begin{equation}
g_{\mu \nu }=\left(
\begin{array}{cccc}
(1-r^{2}/D^{2}) & 0 & 0 & 0 \\
0 & -(1-r^{2}/D^{2})^{-1} & 0 & 0 \\
0 & 0 & -r^{2} & 0 \\
0 & 0 & 0 & -r^{2}\sin ^{2}\theta
\end{array}
\right) .
\end{equation}
Its inverse is
\begin{equation}
g^{\mu \nu }=\left(
\begin{array}{cccc}
-\frac{D^{2}}{-D^{2}+r^{2}} & 0 & 0 & 0 \\
0 & \frac{-D^{2}+r^{2}}{D^{2}} & 0 & 0 \\
0 & 0 & -\frac{1}{r^{2}} & 0 \\
0 & 0 & 0 & -\frac{1}{r^{2}\sin ^{2}\theta }
\end{array}
\right)
\end{equation}
Again proceeding on the same lines for the momentum four vector, we have
\begin{equation}
g=\det (g_{ij})=-r^{4}\sin ^{2}\theta ,\,\,\,f=\frac{1}{\sqrt{g_{00}}}=\frac{%
1}{\sqrt{(1-r^{2}/D^{2})}}\,,\,
\end{equation}
\begin{equation}
\ln \sqrt{-g}=\ln (r^{2}\sin \theta
),\,\,\,\,\,\,\,\,\,\,\,\,\,\,\,\,\,\,\,\,\,\,\,\,\,\,\,\,\,\,\,\,\,\,\,\,\,%
\,\,\,\,\,\,\,\,\,\,\,\,\,\,\,\,\,\,\,\,\,\,\,\,\,\,\,\,\,\,\,\,\,\,
\end{equation}
which implies that

\begin{equation}
Af=(\ln \sqrt{-g})_{,0}=0,\,\,\,\,\,\,\,\,\,\,\,\,\,\,\,\,\,\,\,\,\,\,\,\,\,%
\,\,\,\,\,\,\,\,\,\,\,\,\,\,\,\,\,\,\,\,\,\,\,\,\,\,\,\,\,\,\,\,
\end{equation}
\begin{equation}
(\ln
(Af))_{,0}=0,\,\,\,\,\,\,\,\,\,\,\,\,\,\,\,\,\,\,\,\,\,\,\,\,\,\,\,\,\,\,\,%
\,\,\,\,\,\,\,\,\,\,\,\,\,\,\,\,\,\,\,\,\,\,\,  \label{des1}
\end{equation}
\begin{equation}
g_{,0}^{ij}g_{ij,o}=0.\,\,\,\,\,\,\,\,\,\,\,\,\,\,\,\,\,\,\,\,\,\,\,\,\,\,\,%
\,\,\,\,\,\,\,\,\,\,\,\,\,\,\,\,\,\,\,\,\,\,\,\,\,\,\,  \label{des2}
\end{equation}
So from equations (\ref{des1}), (\ref{des2}) we get

\begin{equation}
F_{0}=0.\,\,\,\,\,\,\,\,\,\,\,\,\,\,\,\,\,\,\,\,\,\,\,\,\,\,\,\,\,\,\,\,\,\,%
\,\,\,\,\,\,\,\,\,\,\,\,\,\,\,\,\,\,\,\,\,\,\,\,\,\,\,\,\,\,\,\,\,\,\,\,\,\,%
\,\,\,\,\,\,\,\,\,\,\,\,\,\,\,\,\,\,\,\,\,\,\,\,\,\,\,\,\,\,\,\,\,\,\,\,
\end{equation}
Hence from equation (\ref{4.1}) we get
\begin{equation}
S=\textrm{{\it constant.}}
\end{equation}
Clearly, we would need to take $S=0$ here.

{\bf 6. The Lemaitre form of the De Sitter universe}

\thinspace \thinspace \thinspace \thinspace \thinspace \thinspace \thinspace
\thinspace \thinspace \thinspace Consider the empty space solution of the
Einstein field equations with cosmological constant,
\begin{equation}
ds^{2}=dt^{2}-a_{0}^{2}e^{2(\Lambda /3)^{1/2}t}\left[ d\chi ^{2}+\chi
^{2}d\theta ^{2}+\chi ^{2}\sin ^{2}\theta d\varphi ^{2}\right] .  \label{d2}
\end{equation}
The metric tensor is
\begin{equation}
g_{\mu \nu }=\left(
\begin{array}{cccc}
1 & 0 & 0 & 0 \\
0 & -a_{0}^{2}e^{2(\Lambda /3)^{1/2}t} & 0 & 0 \\
0 & 0 & -a_{0}^{2}e^{2(\Lambda /3)^{1/2}t}\chi ^{2} & 0 \\
0 & 0 & 0 & -a_{0}^{2}e^{2(\Lambda /3)^{1/2}t}\chi ^{2}\sin ^{2}\theta
\end{array}
\right) .
\end{equation}
Its inverse is
\begin{equation}
g^{\mu \nu }=\left(
\begin{array}{cccc}
1 & 0 & 0 & 0 \\
0 & -\frac{1}{a_{0}^{2}e^{\frac{2}{3}\sqrt{3}\sqrt{\Lambda }t}} & 0 & 0 \\
0 & 0 & -\frac{1}{a_{0}^{2}e^{\frac{2}{3}\sqrt{3}\sqrt{\Lambda }t}\chi ^{2}}
& 0 \\
0 & 0 & 0 & -\frac{1}{a_{0}^{2}e^{\frac{2}{3}\sqrt{3}\sqrt{\Lambda }t}\chi
^{2}\sin ^{2}\theta }
\end{array}
\right) .
\end{equation}

To find the momentum four vector we need the force four vector. We therefore
calculate it. Now

\begin{equation}
g=\det (g_{ij})=-a_{0}^{6}e^{2\sqrt{3}\sqrt{\Lambda }t)}\chi ^{4}\sin
^{2}\theta ,\,\,\,f=\frac{1}{\sqrt{g_{00}}}=1\,\,,
\end{equation}
\begin{equation}
\ln \sqrt{-g}=3\ln a_{0}+2\ln \chi +\sqrt{3}\sqrt{\Lambda }t+\frac{1}{2}\ln
\left( -\cos \theta +1\right) +\frac{1}{2}\ln \left( \cos \theta +1\right) ,
\end{equation}
which implies that

\begin{equation}
Af=(\ln \sqrt{-g})_{,0}=\sqrt{3}\sqrt{\Lambda },\,\,\,\,\,\,\,\,\,\,\,\,\,\,%
\,\,\,\,\,\,\,\,\,\,\,\,\,\,\,\,\,\,\,\,\,\,\,\,\,\,\,\,\,\,\,\,\,\,\,\,\,\,%
\,\,
\end{equation}

\begin{equation}
(\ln
(Af))_{,0}=0,\,\,\,\,\,\,\,\,\,\,\,\,\,\,\,\,\,\,\,\,\,\,\,\,\,\,\,\,\,\,\,%
\,\,\,\,\,\,\,\,\,\,\,\,\,\,\,\,\,\,\,\,\,\,\,  \label{lem1}
\end{equation}

\begin{equation}
g_{,0}^{ij}g_{ij,o}=-4\Lambda
.\,\,\,\,\,\,\,\,\,\,\,\,\,\,\,\,\,\,\,\,\,\,\,\,\,\,\,\,\,\,\,\,
\label{lem2}
\end{equation}
So from equations (\ref{lem1}), (\ref{lem2}) and (\ref{00.5c}) we get
\begin{equation}
F_{0}=m\left( \frac{\Lambda }{3}\right)
^{1/2}.\,\,\,\,\,\,\,\,\,\,\,\,\,\,\,\,\,\,\,\,\,\,\,\,\,\,\,\,\,\,\,\,\,\,\,
\end{equation}
Hence from equation (\ref{4.1}) we get
\begin{equation}
S=m\lambda \left( \frac{\Lambda }{3}\right) ^{1/2}t+\textrm{{\it
constant.}} \label{Lp0}
\end{equation}
This does not seem reasonable. Since the other form of the metric gives a
different result, it is clear that the interpretation is not even internally
consistent.

\section{ The geodesic analysis for angular momentum imparted to test
particles by gravitational waves}

In section $4.1$ we have concluded that $S$ is not the spin angular momentum
imparted to test particles by gravitational waves. Then the question arises,
if $S$ is not the spin then what is the spin angular momentum. So in this
section we use the geodesic analysis $\left[ 4\right] $ to find the angular
momentum imparted to test particles by gravitational waves. This formula is
further applied to various cases.

Consider a time like congruence of the world lines (not necessarily
geodetic) with tangent vector $u^{a}.$ Decompose $u_{a;b}$ by means of the
operator $h_{ab}$ projecting in to the infinitesimal 3-space orthogonal to $%
u^{a}$ $[4].$

\begin{equation}
u_{a;b}=-\omega _{ab}+\sigma _{ab}+\frac{1}{3}\theta h_{ab}-u_{a}^{.}u_{b}.
\end{equation}
where
\begin{equation}
\,-\omega _{ab}\equiv
u_{[a,b]}+u_{[a}^{.}u_{b]}\,,\,\,\,\,\,\,\,\,\,\,\,\,\,\,\,u_{a}^{.}\equiv
u_{a;b}u^{b}.  \label{4.7}
\end{equation}
\begin{equation}
\theta \equiv
u_{;a}^{a}\,\,\,\,,\,\,\,\,\,\,\,\,\,\,\,\,\,\,\,\,\,\,\,\,\,\,\,\,h_{ab}%
\equiv g_{ab}+u_{a}u_{b}
\end{equation}
\begin{equation}
\,\,\,\,\,\,\,\,\,\sigma _{ab}\equiv u_{(a;b)}+u_{(a}^{.}u_{b)}-\frac{1}{3}%
\theta h_{ab}\,\,\,\,\,\,\,\,\,\,\,\,\,(\sigma _{a}^{a}=0).
\end{equation}
For an observer along one of the world lines and using Fermi propagated
axes, $\omega _{ab}$ describe velocity of rotation, $\sigma _{ab\textrm{ }}$%
shear and $\theta $ describe expansion of the neighboring free particles.
Since the $\psi N-$formalism uses the fermi-walker frame, it could be
expected that the results of this analysis should be consistent with it. For
our purpose only $\omega _{ab}$ is needed. Choose the coordinates so that
the tangent vector is $u^{a}=f\delta _{0}^{a}.$ In the case $%
g_{00}=1,\,u^{a}=\delta _{0}^{a}.$ If $g_{00}\neq 1,\,u^{a}=\frac{1}{\sqrt{%
g_{00}}}\delta _{0}^{a}\,$Thus, from equation (\ref{4.7}) we have
\begin{equation}
-\omega _{ab}=(u_{a,b}-u_{b,a})/2+(u_{a}^{.}u_{b}-u_{b}^{.}u_{a})/2.
\end{equation}
The first and second component on the right hand side of the above equation
vanishes. Thus we have$\,\,$%
\begin{equation}
-\omega _{ab}=(u_{a}^{.}u_{b}-u_{b}^{.}u_{a})/2.
\end{equation}
Further using equation (\ref{4.7}) we have

\begin{eqnarray}
-2\omega _{ab} &=&\left\{
\begin{array}{c}
d \\
c\,\,\,\,\,a
\end{array}
\right\} u_{d}u^{c}u_{b}-\left\{
\begin{array}{c}
d \\
c\,\,\,\,\,b
\end{array}
\right\} u_{d}u^{c}u_{a} \\
&=&\left\{
\begin{array}{c}
0 \\
0\,\,\,\,\,a
\end{array}
\right\} \delta _{b}^{0}-\left\{
\begin{array}{c}
0 \\
0\,\,\,\,b
\end{array}
\right\} \delta _{a}^{0}.  \nonumber
\end{eqnarray}
Taking $a\,$to be $i$ and $b$ to be zero, we finally obtain the components
of the spin vector:
\begin{equation}
\omega _{i0}=-\frac{1}{2}\left\{
\begin{array}{c}
0 \\
0\,\,\,\,\,\,\,i
\end{array}
\right\} .\,\,\,\,\,\,\,(i=1,2,3)  \label{4.8}
\end{equation}
$\,\,\,\,\,\,\,\,\,\,\,\,\,\,\,\,\,\,\,\,\,\,\,\,\,\,\,\,\,\,\,\,\,\,\,\,\,%
\,\,\,\,\,\,\,\,\,\,\,\,\,\,\,\,\,\,\,\,\,\,\,\,\,\,\,\,\,\,\,\,\,\,\,\,\,\,%
\,\,\,\,\,\,\,\,\,\,\,\,\,\,\,\,\,$

This simple formula appears because $u^{a}=\delta _{0}^{a}\,$only. This
gives the angular momentum imparted to test particles by gravitational
waves. We now apply this formula to different types of gravitational waves.

\section{ Applications}

\subsection{Plane gravitational waves}

For a metric $g_{\mu \nu }$ construct the new metric

\begin{equation}
\bar{g}_{\mu \nu }=\eta _{\mu \nu }+[g_{\mu \nu }-\eta _{\mu \nu }]u(s-s_{0})
\label{newdef}
\end{equation}
where $\eta _{\mu \nu }$ is the Minkowski spacetime and $u(s-s_{0})$ is the
step function defined as:

$u(s-s_{0})=0\,\,\,\,\,\,\,\,\,\,$\thinspace \thinspace when$%
\,\,\,\,\,\,\,\,\,s<s_{0}\,$

$u(s-s_{0})\,\,=1\,\,\,\,\,\,\,\,\,\,\,\,\,\,$when\thinspace $%
\,\,\,\,\,\,\,s>s_{0}\,.$

$.$According to this definition the metric tensor becomes:
\begin{equation}
\bar{g}_{\mu \nu }=\left(
\begin{array}{cccc}
\left[ (e^{2\Omega (\alpha )}-1)u+1\right] & 0 & 0 & 0 \\
0 & (1-e^{2\Omega (\alpha )})u-1 & 0 & 0 \\
0 & 0 & (-\alpha ^{2}e^{2\beta (\alpha )}+1)u-1 & 0 \\
0 & 0 & 0 & (-\alpha ^{2}e^{-2\beta (\alpha )}+1)u-1
\end{array}
\right) .
\end{equation}
where $\alpha =t-x$. The inverse of this metric is

\begin{equation}
\bar{g}_{\mu \nu }=\left(
\begin{array}{cccc}
\frac{1}{(e^{2\Omega (\alpha )}-1)u+1} & 0 & 0 & 0 \\
0 & \frac{1}{-1+(-e^{2\Omega (\alpha )}+1)u} & 0 & 0 \\
0 & 0 & \frac{1}{-1+(-\alpha ^{2}e^{2\beta (\alpha )}+1)u} & 0 \\
0 & 0 & 0 & \frac{1}{-1+(-\alpha ^{2}e^{-2\beta (\alpha )}+1)u}
\end{array}
\right) .
\end{equation}
Now

\begin{equation}
\left\{
\begin{array}{c}
0 \\
0\,\,\,\,\,\,1
\end{array}
\right\} =\frac{1}{2}g^{00}g_{00,1},\,\,\,\,\left\{
\begin{array}{c}
0 \\
0\,\,\,\,\,\,2
\end{array}
\right\} =\frac{1}{2}g^{00}g_{00,2},\,\,\,\left\{
\begin{array}{c}
0 \\
0\,\,\,\,\,\,3
\end{array}
\right\} =\frac{1}{2}g^{00}g_{00,3}.  \label{4.11}
\end{equation}
If we find $g_{00,1},\,g_{00,2}\,$and $g_{00,3}$ then we are done. Since $%
g_{00}=(e^{2\Omega (\alpha )}-1)u+1$.
\begin{eqnarray}
g_{00,1} &=&[(e^{2\Omega (\alpha )}-1)u+1]_{,1} \\
&=&2\Omega ^{\prime }(\alpha )e^{2\Omega (\alpha )}u-(e^{2\Omega (\alpha
)}-1)\frac{\partial s}{\partial x}\delta (s-s_{0}).  \nonumber
\end{eqnarray}
Dividing equation (\ref{7}) by $ds^{2}$ and keeping $t,y$ and $z$ fixed, we
get:
\begin{equation}
\frac{\partial s}{\partial x}=e^{\Omega (\alpha
)}\,.\,\,\,\,\,\,\,\,\,\,\,\,\,\,\,\,\,\,\,\,\,\,\,\,\,\,\,\,\,\,\,\,\,\,\,%
\,\,\,\,\,\,\,\,\,\,\,\,\,\,\,\,\,\,\,\,\,\,\,\,\,\,\,\,\,\,\,\,\,\,\,\,\,\,%
\,\,\,\,\,\,\,\,\,\,\,\,\,\,\,\,\,\,\,\,\,\,\,\,\,\,\,\,\,\,\,\,\,\,\,\,\,\,%
\,\,\,\,
\end{equation}
Consequently we have

\begin{equation}
g_{00,1}=-2\Omega ^{\prime }(\alpha )e^{2\Omega (\alpha
)}u(s-s_{0})+e^{\Omega (\alpha )}(e^{2\Omega (\alpha )}-1)\delta (s-s_{0}).
\label{4.14}
\end{equation}
Further
\begin{equation}
g_{00,2}=-(e^{2\Omega (\alpha )}-1)\frac{\partial s}{\partial y}\delta
(s-s_{0}).  \label{4.15}
\end{equation}
Again dividing equation (\ref{7}) by $ds^{2}$ and keeping $t,x$ and $z$
fixed, we get
\begin{equation}
\frac{\partial s}{\partial y}=\alpha e^{\beta (\alpha
)}.\,\,\,\,\,\,\,\,\,\,\,\,\,\,\,\,\,\,\,\,\,\,\,\,\,\,\,\,\,\,\,\,\,\,\,\,%
\,\,\,\,\,\,\,\,\,\,\,\,\,\,\,\,\,\,
\end{equation}
Putting this value of$\frac{\partial s}{\partial y}\,$in to equation (\ref
{4.15}) we get

\begin{equation}
g_{00,2}=-e^{\beta (\alpha )}(e^{2\Omega (\alpha )}-1)\delta (s-s_{0}).
\label{4.16}
\end{equation}
Also,
\begin{equation}
g_{00,3}=-\alpha (e^{2\Omega (\alpha )}-1)\frac{\partial s}{\partial z}%
\delta (s-s_{0}).  \label{4.17}
\end{equation}
As in the previous cases, dividing equation (\ref{7}) by $ds^{2}$ and
keeping $t,x$ and $y$ fixed, we get
\begin{equation}
\frac{\partial s}{\partial z}=\alpha e^{-\beta (\alpha
)}.\,\,\,\,\,\,\,\,\,\,\,\,\,\,\,\,\,\,\,\,\,\,\,\,\,\,\,\,\,\,\,\,\,\,\,\,%
\,\,\,\,\,\,\,\,\,\,\,\,\,\,\,
\end{equation}
Putting $d\frac{\partial s}{\partial z}$ in equation (\ref{4.17})
we get:

\begin{equation}
g_{00,3}=-\alpha e^{-\beta (\alpha )}(e^{2\Omega (\alpha )}-1)\delta
(s-s_{0}).  \label{4.18}
\end{equation}
Using equations (\ref{4.14}), (\ref{4.16}) and (\ref{4.18}) in equation (\ref
{4.11}) we get:

\begin{equation}
\left\{
\begin{array}{c}
0 \\
0\,\,\,\,\,1
\end{array}
\right\} =\frac{1}{2(e^{2\Omega (\alpha )}-1)u+1}\left[ -2\Omega ^{\prime
}(\alpha )e^{2\Omega (\alpha )}u(s-s_{0})+e^{\Omega (\alpha )}(e^{2\Omega
(\alpha )}-1)\delta (s-s_{0})\right]  \label{4.19}
\end{equation}
\begin{equation}
\left\{
\begin{array}{c}
0 \\
0\,\,\,\,\,2
\end{array}
\right\} =\frac{1}{2(e^{2\Omega (\alpha )}-1)u+1}\left[ \alpha e^{\beta
(\alpha )}(e^{2\Omega (\alpha )}-1)\delta (s-s_{0})\right]
,\,\,\,\,\,\,\,\,\,\,\,\,\,\,\,\,\,\,\,\,\,\,\,\,\,\,\,\,\,\,\,\,\,\,
\label{4.20}
\end{equation}
\begin{equation}
\left\{
\begin{array}{c}
0 \\
0\,\,\,\,\,3
\end{array}
\right\} =\frac{1}{2(e^{2\Omega (\alpha )}-1)u+1}\left[ \alpha e^{-\beta
(\alpha )}(e^{2\Omega (\alpha )}-1)\delta (s-s_{0})\right]
.\,\,\,\,\,\,\,\,\,\,\,\,\,\,\,\,\,\,\,\,\,\,\,\,\,\,\,\,\,\,\,\,\,
\label{4.21}
\end{equation}
Now using equations (\ref{4.19}), (\ref{4.20}) and (\ref{4.21}) in equation (%
\ref{4.8}) we get components of the spin vector

\begin{equation}
\omega _{10}=-\frac{1}{4(e^{2\Omega (\alpha )}-1)u+1}\left[ -2\Omega
^{\prime }(\alpha )e^{2\Omega (\alpha )}u(s-s_{0})+e^{\Omega (\alpha
)}(e^{2\Omega (\alpha )}-1)\delta (s-s_{0})\right] ,  \label{4.86}
\end{equation}
\begin{equation}
\omega _{20}=-\frac{1}{4(e^{2\Omega (\alpha )}-1)u+1}\left[ \alpha e^{\beta
(\alpha )}(e^{2\Omega (\alpha )}-1)\delta (s-s_{0})\right]
,\,\,\,\,\,\,\,\,\,\,\,\,\,\,\,\,\,\,\,\,\,\,\,\,\,\,\,\,\,\,\,\,\,\,\,\,\,%
\,\,\,\,\,\,\,\,\,\,\,\,\,\,
\end{equation}
\begin{equation}
\omega _{30}=-\frac{1}{4(e^{2\Omega (\alpha )}-1)u+1}\left[ \alpha e^{-\beta
(\alpha )}(e^{2\Omega (\alpha )}-1)\delta (s-s_{0})\right]
.\,\,\,\,\,\,\,\,\,\,\,\,\,\,\,\,\,\,\,\,\,\,\,\,\,\,\,\,\,\,\,\,\,\,\,\,\,%
\,\,\,\,\,\,\,
\end{equation}
In equation (\ref{4.86}) the first part in the brackets gives the spin
imparted to test particles by the plane gravitational waves while the second
part gives the spin of the wave itself.

\subsection{Cylindrical gravitational waves}

Now use the same construction for the metric (\ref{2.1}) to get

\begin{eqnarray}
g_{00} &=&(e^{2(\gamma -\psi )}-1)u(s-s_{0})+1,  \nonumber \\
g_{11} &=&-(e^{2(\gamma -\psi )}-1)u(s-s_{0})-1,  \nonumber \\
g_{22} &=&-\rho ^{2}\left[ (e^{-2\psi }-1)u(s-s_{0})+1\right] ,  \nonumber \\
g_{33} &=&-(e^{2\psi }-1)u(s-s_{0})+1.
\end{eqnarray}
The inverse of this metric is
\begin{equation}
\bar{g}^{\mu \nu }=\left(
\begin{array}{llll}
g^{00} & 0 & 0 & 0 \\
0 & g^{11} & 0 & 0 \\
0 & 0 & g^{22} & 0 \\
0 & 0 & 0 & g^{33}
\end{array}
\right) ,
\end{equation}
where

$g^{00}=\frac{1}{(e^{2(\gamma -\psi )}-1)u(s-s_{0})+1},$

$g^{11}=\frac{1}{-(e^{2(\gamma -\psi )}-1)u(s-s_{0})-1},$

$g^{22}=\frac{1}{-\rho ^{2}\left[ (e^{-2\psi }-1)u(s-s_{0})+1\right] },$

$g^{33}=\frac{1}{-(e^{2\psi }-1)u(s-s_{0})-1}.$

In this case

\begin{equation}
g_{00,1}=2e^{2\left( \gamma -\psi \right) }(\gamma ^{\prime }-\psi ^{\prime
})u(s-s_{0})+(e^{2\left( \gamma -\psi \right) }-1)\frac{\partial s}{\partial
\rho }\delta (s-s_{0)}.  \label{c1}
\end{equation}
Dividing equation (\ref{2.1}) by $ds^{2}$ and keeping $t,\varphi ,z$ fixed,
we get

\begin{equation}
1=e^{2\left( \gamma -\psi \right) }\left( \frac{\partial \rho }{\partial s}%
\right)
^{2},\,\,\,\,\,\,\,\,\,\,\,\,\,\,\,\,\,\,\,\,\,\,\,\,\,\,\,\,\,\,\,\,\,\,\,%
\,\,\,\,\,\,\,\,\,\,\,\,\,\,\,\,\,
\end{equation}
which implies that

\begin{equation}
\frac{\partial s}{\partial \rho }=e^{\left( \gamma -\psi \right)
}.\,\,\,\,\,\,\,\,\,\,\,\,\,\,\,\,\,\,\,\,\,\,\,\,\,\,\,\,\,\,\,\,\,\,\,\,\,%
\,\,\,\,\,\,\,\,\,\,\,\,\,\,\,\,\,\,\,\,\,\,\,\,\,\,\,\,\,\,\,\,\,\,
\label{qwe}
\end{equation}
Now putting $\frac{\partial s}{\partial \rho }$ in equation (\ref{c1}) we get

\begin{equation}
g_{00,1}=+2e^{2\left( \gamma -\psi \right) }\left( \gamma ^{^{\prime }}-\psi
^{^{\prime }}\right) u\left( s-s_{0}\right) +e^{\gamma -\psi }\left(
e^{2\left( \gamma -\psi \right) }-1\right) \delta \left( s-s_{0}\right) ,
\label{C2}
\end{equation}
\begin{eqnarray}
g_{00,2} &=&-\left[ \left( e^{2\left( \gamma -\psi \right) }-1\right) u+1%
\right] _{,2}\,\,\,\,\,\,\,\,\,\,\,\,\,\,\,\,\,\,\,\,\,\,\,\,\,\,\,\,\,\,\,%
\,\,\,\,\,\,\,\,\,\,\,\,\,\,\,\,\,\,\,\,\,\,\,\,\,\,\,\,\,\,\,\,\,\,\,\,\,\,%
\,\,\,\,\,\,\,\,\,\,\,\,\,\,\,\,\,\,\,\,\,\,\,\,\,\,\,\,\,\,\,\,\,  \nonumber
\\
&=&-(e^{2\left( \gamma -\psi \right) }-1)\frac{\partial s}{\partial \varphi }%
\delta \left( s-s_{0}\right) ,  \label{c3}
\end{eqnarray}
\begin{equation}
g_{00,3}=-\left( e^{2\left( \gamma -\psi \right) }-1\right) \frac{\partial s%
}{\partial z}\delta \left( s-s_{0}\right)
.\,\,\,\,\,\,\,\,\,\,\,\,\,\,\,\,\,\,\,\,\,\,\,\,\,\,\,\,\,\,\,\,\,\,\,\,\,%
\,\,\,\,\,\,\,\,\,\,\,\,\,\,\,\,\,\,\,\,\,\,\,\,\,\,\,\,\,\,\,\,\,\,\,\,\,\,%
\,\,\,\,\,\,\,\,\,\,\,\,\,\,\,\,\,\,\,\,\,\,\,\,\,\,\,\,\,\,  \label{c4}
\end{equation}
Dividing equation (\ref{2.1}) by $ds^{2}$ and keeping $t$,$\rho ,z$ fixed,
we get

\begin{equation}
1=e^{2\psi }\rho ^{2}\left( \frac{\partial y}{\partial s}\right)
^{2},\,\,\,\,\,\,\,\,\,\,\,\,\,\,\,\,\,\,\,\,\,\,\,\,\,\,\,\,\,\,\,\,\,\,\,%
\,\,\,\,\,\,\,\,\,\,\,\,\,\,\,\,\,\,\,\,\,\,\,\,\,\,\,\,\,\,\,\,\,\,\,\,\,\,%
\,\,\,\,\,\,\,\,\,\,\,\,\,\,\,\,\,\,\,\,\,\,\,\,\,\,\,\,\,\,\,\,\,\,\,\,\,\,%
\,\,\,\,\,\,\,
\end{equation}
which implies that

\begin{equation}
\frac{\partial s}{\partial y}=\rho e^{-\psi
}.\,\,\,\,\,\,\,\,\,\,\,\,\,\,\,\,\,\,\,\,\,\,\,\,\,\,\,\,\,\,\,\,\,\,\,\,\,%
\,\,\,\,\,\,\,\,\,\,\,\,\,\,\,\,\,\,\,\,\,\,\,\,\,\,\,\,\,\,\,\,\,\,\,\,\,\,%
\,\,\,\,\,\,\,\,\,\,\,\,\,\,\,\,\,\,\,\,\,\,\,\,\,\,\,\,\,\,\,\,\,\,\,\,\,\,%
\,\,\,\,\,\,\,\,\,\,\,\,\,\,\,\,\,\,\,\,\,\,\,\,\,\,\,  \label{c5}
\end{equation}
Now dividing equation (\ref{2.1}) by $ds^{2}$and keeping $t,\rho ,y$ fixed,
we get
\begin{equation}
1=e^{2\psi }\left( \frac{\partial z}{\partial s}\right)
^{2}\,\,\,\,\,\,\,\,\,\,\,\,\,\,\,\,\,\,\,\,\,\,\,\,\,\,\,\,\,\,\,\,\,\,\,\,%
\,\,\,\,\,\,\,\,\,\,\,\,\,\,\,\,\,\,\,\,\,\,\,\,\,\,\,\,\,\,\,\,\,\,\,\,\,\,%
\,\,\,\,\,\,\,\,\,\,\,\,\,\,\,\,\,\,\,\,\,\,\,\,\,\,\,\,\,\,\,\,\,\,\,\,\,\,%
\,\,\,\,\,\,\,\,\,\,\,\,\,\,\,\,\,\,
\end{equation}
\begin{equation}
\frac{\partial s}{\partial z}=e^{\psi
}.\,\,\,\,\,\,\,\,\,\,\,\,\,\,\,\,\,\,\,\,\,\,\,\,\,\,\,\,\,\,\,\,\,\,\,\,\,%
\,\,\,\,\,\,\,\,\,\,\,\,\,\,\,\,\,\,\,\,\,\,\,\,\,\,\,\,\,\,\,\,\,\,\,\,\,\,%
\,\,\,\,\,\,\,\,\,\,\,\,\,\,\,\,\,\,\,\,\,\,\,\,\,\,\,\,\,\,\,\,\,\,\,\,\,\,%
\,\,\,\,\,\,\,\,\,\,\,\,\,\,\,\,\,\,\,\,\,\,\,\,\,\,\,\,\,\,\,\,\,\,\,
\label{c6}
\end{equation}
Putting equation (\ref{c5}) in to equation (\ref{c3}) we get,
\begin{equation}
g_{00,2}=-\left( e^{2\left( \gamma -\psi \right) }-1\right) \rho e^{-\psi
}\delta \left( s-s_{0}\right)
.\,\,\,\,\,\,\,\,\,\,\,\,\,\,\,\,\,\,\,\,\,\,\,\,\,\,\,\,\,\,\,\,\,\,\,\,\,%
\,\,\,\,\,\,\,\,\,\,\,\,\,\,\,\,\,\,\,\,\,\,\,\,\,\,\,\,\,\,\,\,\,\,\,\,\,\,%
\,\,\,\,\,\,\,\,\,\,\,\,\,\,\,\,\,\,\,\,\,\,\,\,\,\,\,\,\,\,\,  \label{c7}
\end{equation}
Putting equation (\ref{c6}) in to equation (\ref{c4}) we get,
\begin{equation}
g_{00,3}=-\left( e^{2\left( \gamma -\psi \right) }-1\right) e^{\psi }\delta
\left( s-s_{0}\right)
.\,\,\,\,\,\,\,\,\,\,\,\,\,\,\,\,\,\,\,\,\,\,\,\,\,\,\,\,\,\,\,\,\,\,\,\,\,%
\,\,\,\,\,\,\,\,\,\,\,\,\,\,\,\,\,\,\,\,\,\,\,\,\,\,\,\,\,\,\,\,\,\,\,\,\,\,%
\,\,\,\,\,\,\,\,\,\,\,\,\,\,\,\,\,\,\,\,\,\,\,\,\,\,\,\,\,\,\,\,\,\,
\label{c8}
\end{equation}
Using equations (\ref{C2}), (\ref{c7}) and (\ref{c8}) in equation (\ref{4.11}%
) we get:
\begin{equation}
\left\{
\begin{array}{c}
0 \\
0\,\,\,\,\,1
\end{array}
\right\} =-\frac{1}{2(e^{2(\gamma -\psi )}-1)u+1}\left[ 2e^{2(\gamma -\psi
)}(\gamma ^{\prime }-\psi ^{\prime })u+e^{\gamma -\psi }(e^{2(\gamma -\psi
)}-1)\delta (s-s_{0})\right] ,  \label{c9}
\end{equation}
\begin{equation}
\left\{
\begin{array}{c}
0 \\
0\,\,\,\,\,2
\end{array}
\right\} =-\frac{1}{2(c^{2}e^{2(\gamma -\psi )}-1)u+1}\left[ (e^{2(\gamma
-\psi )}-1)\rho e^{-\psi }\delta (s-s_{0})\right] ,\,\,\,\,\,\,\,\,\,\,\,\,%
\,\,\,\,\,\,\,\,\,\,\,\,\,\,\,\,\,\,\,\,\,\,\,\,\,\,\,\,\,\,\,\,\,\,\,\,\,\,%
\,\,\,\,\,\,\,\,\,  \label{c10}
\end{equation}
\begin{equation}
\left\{
\begin{array}{c}
0 \\
0\,\,\,\,\,3
\end{array}
\right\} =-\frac{1}{2(e^{2(\gamma -\psi )}-1)u+1}\left[ (c^{2}e^{2(\gamma
-\psi )}-1)e^{\psi }\delta (s-s_{0})\right] .\,\,\,\,\,\,\,\,\,\,\,\,\,\,\,%
\,\,\,\,\,\,\,\,\,\,\,\,\,\,\,\,\,\,\,\,\,\,\,\,\,\,\,\,\,\,\,\,\,\,\,\,\,\,%
\,\,\,\,\,\,\,\,\,\,  \label{c11}
\end{equation}
Now using equations (\ref{c9}), (\ref{c10}) and (\ref{c11}) in equation (\ref
{4.8}) we get the components of the spin vector

\begin{equation}
\omega _{10}=\frac{1}{4(e^{2(\gamma -\psi )}-1)u+1}\left[ 2e^{2(\gamma -\psi
)}(\gamma ^{\prime }-\psi ^{\prime })u+e^{\gamma -\psi }(e^{2(\gamma -\psi
)}-1)\delta (s-s_{0})\right] ,
\end{equation}
\begin{equation}
\omega _{20}=\frac{1}{4(e^{2(\gamma -\psi )}-1)u+1}\left[ (e^{2(\gamma -\psi
)}-1)\rho e^{-\psi }\delta (s-s_{0})\right] ,\,\,\,\,\,\,\,\,\,\,\,\,\,\,\,%
\,\,\,\,\,\,\,\,\,\,\,\,\,\,\,\,\,\,\,\,\,\,\,\,\,\,\,\,\,\,\,\,\,\,\,\,
\end{equation}
\begin{equation}
\omega _{30}=\frac{1}{4(e^{2(\gamma -\psi )}-1)u+1}\left[ (e^{2(\gamma -\psi
)}-1)e^{\psi }\delta (s-s_{0})\right] .\,\,\,\,\,\,\,\,\,\,\,\,\,\,\,\,\,\,%
\,\,\,\,\,\,\,\,\,\,\,\,\,\,\,\,\,\,\,\,\,\,\,\,\,\,\,\,\,\,\,\,\,\,\,\,\,\,
\end{equation}

\subsection{ The Friedmann model}

Consider the Friedmann model. The metric for this model is given by equation
(\ref{4.2}) and the metric tensor is
\begin{equation}
g_{\mu \nu }=\left(
\begin{array}{cccc}
1 & 0 & 0 & 0 \\
0 & -a^{2}(t) & 0 & 0 \\
0 & 0 & -a^{2}(t)f_{k}^{2}(\chi ) & 0 \\
0 & 0 & 0 & -a^{2}(t)f_{k}^{2}(\chi )\sin ^{2}\theta
\end{array}
\right) .
\end{equation}
According to equation (\ref{4.8}) the only thing we need is $%
g_{00,1},\,g_{00,2},\,$and $\,g_{00,3}.$ As there are no wave fronts.
Therefore we have $g_{00}=1.$ Here obviously all it's derivatives will
vanish. Hence
\begin{equation}
\left\{
\begin{array}{c}
0 \\
0\,\,\,\,\,1
\end{array}
\right\} =\left\{
\begin{array}{c}
0 \\
0\,\,\,\,\,2
\end{array}
\right\} =\left\{
\begin{array}{c}
0 \\
0\,\,\,\,\,3
\end{array}
\right\} =0.
\end{equation}
Thus all components of spin vector are zero. i.e.
\begin{equation}
\omega _{10}=\omega _{20}=\omega _{30}=0.
\end{equation}
As we have stated earlier, there is no spin in an isotropic and homogeneous
universe model, this analysis gives zero spin for this model as required.

\subsection{ The Kasner model}

Consider the Kasner model. The metric for the Kasner model is given by
equation (\ref{k}) and the metric tensor is
\begin{equation}
g_{\mu \nu }=\left(
\begin{array}{cccc}
1 & 0 & 0 & 0 \\
0 & -t^{2p_{1}} & 0 & 0 \\
0 & 0 & -t^{2p_{2}} & 0 \\
0 & 0 & 0 & -t^{2p3}
\end{array}
\right) .  \label{K}
\end{equation}
According to equation (\ref{4.8}) the only thing we need is $%
g_{00,1},\,g_{00,2},\,$and $\,g_{00,3}.$ As there is no ``wave-front'' so
the original metric stands. From (\ref{K}) we have $g_{00}=1.$ Here
obviously all it's derivatives will vanish. Hence
\begin{equation}
\left\{
\begin{array}{c}
0 \\
0\,\,\,\,\,1
\end{array}
\right\} =\left\{
\begin{array}{c}
0 \\
0\,\,\,\,\,2
\end{array}
\right\} =\left\{
\begin{array}{c}
0 \\
0\,\,\,\,\,3
\end{array}
\right\} =0.
\end{equation}
Thus all components of spin vector are zero. i.e.
\begin{equation}
\omega _{10}=\omega _{20}=\omega _{30}=0.
\end{equation}

\subsection{The De Sitter universe (usual coordinates)}

The metric for this model is given by equation (\ref{des}) and the metric
tensor is
\begin{equation}
g_{\mu \nu }=\left(
\begin{array}{cccc}
(1-r^{2}/D^{2}) & 0 & 0 & 0 \\
0 & -(1-r^{2}/D^{2})^{-1} & 0 & 0 \\
0 & 0 & -r^{2} & 0 \\
0 & 0 & 0 & -r^{2}\sin ^{2}\theta
\end{array}
\right) .  \label{4.117}
\end{equation}
As $g_{00}\neq 1$ so $\omega _{i0}$ calculated in section $4.2$ cannot work.
Let $u^{a}=f\delta _{0}^{a}.$ Here by definition
\begin{equation}
u^{a}u^{b}g_{ab}=-1\,=(u^{0})^{2}g_{00}\textrm{\thinspace or }\,u^{0}=\frac{1}{%
\sqrt{g_{00}}}
\end{equation}

consequently we have
\begin{equation}
u^{a}=\frac{1}{\sqrt{g_{00}}}\delta _{0}^{a}.
\end{equation}
Now we only need to calculate $u_{[a,b]}\,$and\thinspace $\,u_{[a}^{.}u_{b]}$%
\begin{eqnarray}
u_{a,b}\, &=&\left( \sqrt{g_{00}}\delta _{a}^{0}\right) _{,b}  \nonumber \\
&=&(\ln \sqrt{g_{00}})_{,1}\delta _{a}^{0}\delta _{b}^{1}.
\end{eqnarray}

Hence we have
\begin{equation}
u_{\left[ a,b\right] }=(\ln \sqrt{g_{00}})_{,1}\delta _{[a}^{0}\delta
_{b]}^{1}  \label{4.65}
\end{equation}
Similarly we get
\begin{equation}
u_{[a}^{.}u_{b]}=-\frac{1}{2}\left( \delta _{b}^{0}\delta _{a}^{1}-\delta
_{a}^{0}\delta _{b}^{1}\right) \sqrt{g_{00}}\left\{
\begin{array}{c}
0 \\
0\,\,\,\,1
\end{array}
\right\}  \label{4.66}
\end{equation}
From equation $\left( \ref{4.65}\right) $ and $\left( \ref{4.66}\right) $
the sum $u_{\left[ a,b\right] }+u_{[a}^{.}u_{b]}$ for this metric vanishes
and thus equation $\left( \ref{4.7}\right) $ implies that $\omega _{ab}$
vanishes.

\subsection{The Lemaitre form of the De Sitter universe}

The metric for this model is given by (\ref{d2}). The metric tensor is
\begin{equation}
g_{\mu \nu }=\left(
\begin{array}{cccc}
1 & 0 & 0 & 0 \\
0 & -a_{0}^{2}e^{2(\Lambda /3)^{1/2}t} & 0 & 0 \\
0 & 0 & -a_{0}^{2}e^{2(\Lambda /3)^{1/2}t}\chi ^{2} & 0 \\
0 & 0 & 0 & -a_{0}^{2}e^{2(\Lambda /3)^{1/2}t}\chi ^{2}\sin ^{2}\theta
\end{array}
\right) .
\end{equation}
By the definition (\ref{newdef}) of the metric tensor $g_{00}=1.\,$Here
obviously all it's derivatives will vanish. Hence
\begin{equation}
\left\{
\begin{array}{c}
0 \\
0\,\,\,\,\,1
\end{array}
\right\} =\left\{
\begin{array}{c}
0 \\
0\,\,\,\,\,2
\end{array}
\right\} =\left\{
\begin{array}{c}
0 \\
0\,\,\,\,\,3
\end{array}
\right\} =0.
\end{equation}
Thus all components of spin vector are zero. i.e.
\begin{equation}
\omega _{10}=\omega _{20}=\omega _{30}=0.
\end{equation}
Thus we have got the same result for both the forms of the De Sitter
universe which proves that the formalism is consistent.

\subsection{The  G $\stackrel{..}{o}$del universe model}

Consider the G$\stackrel{..}{o}$del (spinning) universe model
\begin{equation}
ds^{2}=dt^{2}-dx^{2}-dy^{2}+2m(x)dtdz-l(x)dz^{2},  \label{G}
\end{equation}
where $m(x)=Ae^{ax}$ and $l(x)=A^{2}(\frac{a^{2}}{b^{2}}-1)e^{2ax}.\,$The
metric tensor is
\begin{equation}
g_{\mu \nu }=\left(
\begin{array}{cccc}
1 & 0 & 0 & m(x) \\
0 & -1 & 0 & 0 \\
0 & 0 & -1 & 0 \\
m(x) & 0 & 0 & -l(x)
\end{array}
\right) ,
\end{equation}
Its inverse is
\begin{equation}
g^{\mu \nu }=\left(
\begin{array}{cccc}
\frac{l(x)}{l(x)+m^{2}(x)} & 0 & 0 & \frac{m(x)}{l(x)+m^{2}(x)} \\
0 & -1 & 0 & 0 \\
0 & 0 & -1 & 0 \\
\frac{m(x)}{l(x)+m^{2}(x)} & 0 & 0 & -\frac{1}{l(x)+m^{2}(x)}
\end{array}
\right)
\end{equation}
According to equation (\ref{4.8}) the only thing we need are the following
three Christoffel symbols$\,\left\{
\begin{array}{c}
0 \\
0\,\,\,\,\,1
\end{array}
\right\} $, $\left\{
\begin{array}{c}
0 \\
0\,\,\,\,\,2
\end{array}
\right\} $ and $\left\{
\begin{array}{c}
0 \\
0\,\,\,\,\,3
\end{array}
\right\} .$ Now

\begin{eqnarray}
\left\{
\begin{array}{c}
0 \\
0\,\,\,\,\,1
\end{array}
\right\} &=&\frac{1}{2}g^{03}g_{03,1}  \nonumber \\
&=&\frac{1}{2}\frac{m(x)}{l(x)+m^{2}(x)}(Ae^{ax})_{,1}  \nonumber \\
&=&\frac{1}{2}\frac{am^{2}(x)}{l(x)+m^{2}(x)},
\end{eqnarray}
\begin{equation}
\left\{
\begin{array}{c}
0 \\
0\,\,\,\,\,2
\end{array}
\right\} =\left\{
\begin{array}{c}
0 \\
0\,\,\,\,\,3
\end{array}
\right\} =0.\,\,\,\,\,\,\,\,\,\,\,\,\,\,\,\,
\end{equation}
Thus all components of the spin vector are
\begin{eqnarray}
\omega _{10} &=&-\frac{1}{4}\frac{am^{2}(x)}{l(x)+m^{2}(x)}  \nonumber \\
\omega _{20} &=&\omega _{30}=0.
\end{eqnarray}

\chapter{SUMMARY AND CONCLUSION}

Linearized General Relativity predicts gravitational waves. These
waves are analogous to electromagnetic waves. There may be
different types of gravitational waves i.e. plane and cylindrical
gravitational waves etc. This background was reviewed in chapter
1. Work on plane and cylindrical gravitational waves $\left[
3\right] $ has been helpful in understanding them further. The
question of the reality of gravitational waves was discussed in
chapter 2. There we used Weber-Wheeler $\left[ 3\right] $ method
for a particle in the path of a plane gravitational wave and
obtained the standard constant momentum imparted to the particle.
Some astrophysical sources of gravitational waves were also
discussed. In chapter 3 the work of Qadir and Sharif $\left[
8\right] $ in which a general formula is developed for the
momentum imparted to test particles by gravitational waves in
arbitrary spacetime, was reviewed. In this paper the problem of
the identification of their zero component of the momentum four
vector was mentioned. Sharif $\left[ 11\right] $ gave a suggestion
that it could be interpreted as the spin imparted to a test rod in
an arbitrary spacetime. This suggestion was reviewed in chapter 4.
Further analysis by Sharif $\left[ 14\right] $ was shown here to
provide a counterexample for this suggestion i.e. it gives a non
zero spin for an isotropic and homogeneous universe model.
Further, when the example of the De Sitter universe was
considered, the interpretation gave different results for the
static and Lemaitre forms. This proved that the interpretation is
not even internally consistent. As such $P_{0}$ cannot be
interpreted as the spin imparted to test rods. A geodesic analysis
$\left[ 4\right] $ was used in section $4.2$ to evaluate the spin
imparted to test particles in various cases. In the cases of plane
and cylindrical gravitational waves we got very reasonable
results. As required , the Friedmann, the Kasner and De Sitter
models do not impart spin to test particles according to this
analysis. Further we got the same results for both forms of the De
Sitter universe, which confirms the validity of our analysis.
Finally, the G$\stackrel{..}{o}$del universe model gave a non-zero
$\left( \textrm{non-constant}\right) $ angular momentum, precisely
as it should.

If $P_{0}$ does not provide an expression for the spin imparted to test rods
in an arbitrary spacetime then what is its correct interpretation?. Consider
$E=\sqrt{P^{i}P^{j}g_{ij}+m^{2}}$ and $P_{0}=\int F_{0}dt.$ Define the
difference between the two as $\Delta E=P^{0}-E$ for some of the spacetimes
considered earlier. By definition the usual energy $E$ associated with a
momentum $P^{i}$ is given by
\begin{equation}
P^{\mu }P^{\nu }g_{\mu \nu }=m^{2}=\left( E^{2}\right)
g_{00}+P^{i}P^{j}g_{ij}.
\end{equation}
Hence
\begin{equation}
E=\frac{1}{\sqrt{g_{00}}}\left( m^{2}-P^{i}P^{j}g_{ij}\right) ^{1/2}.
\end{equation}
Now, we want to compare this with
\begin{equation}
P^{0}=g^{00}\int F_{0}dt.
\end{equation}
The energy difference, $\Delta E,$ is then
\begin{equation}
\Delta E=g^{00}\int F_{0}dt-\frac{1}{\sqrt{g_{00}}}\left(
m^{2}-P^{i}P^{j}g_{ij}\right) ^{1/2}.
\end{equation}
When $F_{0}$ is zero we get no further understanding from $\Delta E$. As
such we will not compute these cases. Again for G$\stackrel{..}{o}$del
universe we have not calculated $F_{0}$ as the metric is not in the block
diagonalized form $\left( g_{0i}=0\right) \,$and a gauge transformation
would need to be made before it could be applied. Therefore we only work out
$\Delta E$ for the cases $\left( a\right) $ cylindrical gravitational waves;
$\left( b\right) $ the Friedmann models and $\left( c\right) $ the Lemaitre
form of the De Sitter universe. Finally we will discuss the consequences of
these results.

$\left( {\bf a}\right) $ {\bf Cylindrical gravitational waves}

\smallskip Consider the metric given by equation $\left( \ref{2.1}\right) $
and use the non-zero components of momentum four vector given by equations $%
\left( \ref{00.23}\right) $ and $\left( \ref{00.24}\right) $ to get
\begin{equation}
E=me^{-(\gamma -\psi )}\left( e^{-2\left( \gamma -\psi \right) }\left[
AJ_{0}^{\prime }\sin (\omega t)-\frac{1}{4}A^{2}J_{0}J_{0}^{\prime }+\rho
\omega (J_{0}J_{0}^{\prime })^{\prime }\sin (2\omega t)\right] ^{2}+1\right)
^{1/2}
\end{equation}
Consequently we have

\[
\Delta E=-me^{-(\gamma -\psi )}[(1-e^{-2\left( \gamma -\psi \right)
}(AJ_{0}^{\prime }\sin (\omega t)-\frac{1}{4}A^{2}J_{0}J_{0}^{\prime }+\rho
\omega (J_{0}J_{0}^{\prime })^{\prime }\sin (2\omega t))^{2})^{1/2}
\]
\begin{equation}
\;\;\;\;\;\;\;\;\;\;\;+e^{-\left( \gamma -\psi \right) }(\ln \mid AJ_{0}\sin
(\omega t)\mid +(1+AJ_{0}/\omega \rho J_{0}^{\prime })\ln \mid 1-2\omega
\rho AJ_{0}^{\prime }\cos (\omega t]\mid )].
\end{equation}
\smallskip

$\left( {\bf b}\right) ${\bf The Friedmann models}

Since $P^{1},P^{2}$ and $P^{3}$ are zero for the Friedmann models, from
equations $\left( \ref{4.2}\right) ,$ $\left( \ref{4.3}\right) $, $\left(
\ref{4.4a}\right) $ \smallskip and $\left( \ref{4.5**}\right) $ \smallskip
we get the energy difference for the three models:
\begin{equation}
\Delta E_{k=1}=m[\ln \sqrt{1-\cos \eta }-\frac{3}{8}\cos \eta +\frac{1}{16}%
\cos ^{2}\eta +\ln \sqrt{2}+\frac{5}{16}];
\end{equation}
\begin{equation}
\Delta E_{k=0}=m\ln (\frac{\eta }{2});\,\,\,\,\,\,\,\,\,\,\,\,\,\,\,\,\,\,\,%
\,\,\,\,\,\,\,\,\,\,\,\,\,\,\,\,\,\,\,\,\,\,\,\,\,\,\,\,\,\,\,\,\,\,\,\,\,\,%
\,\,\,\,\,\,\,\,\,\,\,\,\,\,\,\,\,\,\,\,\,\,\,\,\,\,\,\,\,\,\,\,\,\,\,\,\,\,%
\,\,\,\,\,\,\,\,\,\,\,\,\,\,\,\,\,\,\,\,\,\,\,\,\,\,\,
\end{equation}
\begin{equation}
\Delta E_{k=-1}=m\ln \frac{\cosh \eta -1}{\sinh \eta }.\,\,\,\,\,\,\,\,\,\,%
\,\,\,\,\,\,\,\,\,\,\,\,\,\,\,\,\,\,\,\,\,\,\,\,\,\,\,\,\,\,\,\,\,\,\,\,\,\,%
\,\,\,\,\,\,\,\,\,\,\,\,\,\,\,\,\,\,\,\,\,\,\,\,\,\,\,\,\,\,\,\,\,\,\,\,\,\,%
\,\,\,\,\,\,\,\,\,
\end{equation}
$.$

These are plotted in Fig. $\left( 5.1\right) ,\,\left( 5.2\right) \,$and $%
\left( 5.3\right) \,$and discussed shortly.

$\left( {\bf c}\right) ${\bf \ The Lemaitre form of the De Sitter universe}

Again $P^{1},\,P^{2}\,$and $P^{3}$ are zero for the metric given by equation
$\left( \ref{d2}\right) .$ Thus using equation $\left( \ref{Lp0}\right) $ we
get
\begin{equation}
\Delta E=m\sqrt{\Lambda /3}t.\,\,\,\,\,\,\,\,\,\,\,\,\,\,\,\,\,\,\,\,\,\,\,%
\,\,\,\,\,\,\,\,\,\,\,\,\,\,\,\,\,\,\,\,\,\,\,\,\,\,\,\,\,\,\,\,\,\,\,\,\,\,%
\,\,\,\,\,\,\,\,\,\,\,\,\,\,\,\,\,\,\,\,\,\,\,\,\,\,\,\,\,\,\,\,\,\,\,\,\,\,%
\,\,\,\,\,\,\,\,\,\,\,\,\,\,\,\,\,\,\,\,\,\,\,\,\,\,\,\,\,\,\,\,\,\,\,\,\,\,%
\,\,\,\,\,\,\,\,\,\,\,\,\,\,\,\,\,\,\,\,\,\,\,\,\,
\end{equation}
as shown in Fig $\left( 5.4\right) .$

The three expressions for the Friedmann models have the same asymptotic
behavior for sufficiently small values of $\eta $, namely $\Delta E\sim m\ln
\eta $. For $k=\pm 1$ there is a correction term $\sim \pm \frac{\eta ^{2}}{%
12}$. We have inserted a constant term for $k=+1$ so that the expressions
for the three should match up to the zero order terms. At the phase of
maximum expansion of the closed model we get $\Delta E=m(\ln 2+\frac{3}{4}).$
We could equally well, have set $\Delta E=0$ at the phase of maximum
expansion and had a difference for it from the other two cases for small $%
\eta $ (in the constant term). Note that for all the three models $\Delta E$
diverges as $\eta \rightarrow 0$ and it also diverges as $\eta \rightarrow
2\pi $ for the closed model. For $k=+1$ we have chosen to display the
constant term so that $\Delta E=0$ at $\eta =\pi .$ The proposal does not
seem inconsistent, but still needs further discussion.

\newpage

\smallskip

\smallskip

\smallskip

\smallskip

\begin{figure}
\centerline{\epsfig{file=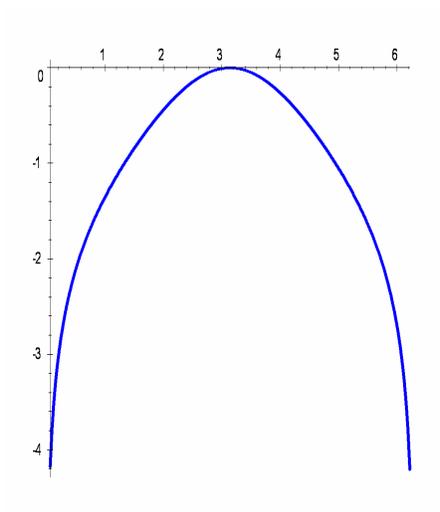,width=6cm,height=7cm}}
\caption{Fig 5.1: $\Delta E$ for the closed friedmann model. This
becomes infinite at the bang and the crunch and becomes zero at
the phase of maximum expansion.} \label{Closed Friedman Model}
\end{figure}
\pagebreak
\begin{figure}
\centerline{\epsfig{file=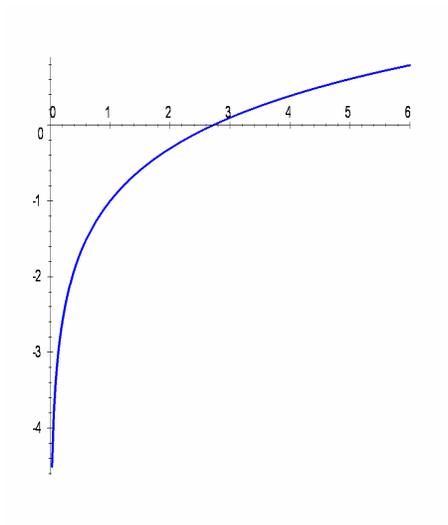,width=6cm,height=7cm}}
\caption{Fig 5.2: $\Delta E$ for the flat Friedmann model, again
$m$ is taken to be unity. $\Delta E$ diverges at $\protect\eta
=0.$} \label{saA}
\end{figure}

\newpage
\begin{figure}
\centerline{\epsfig{file=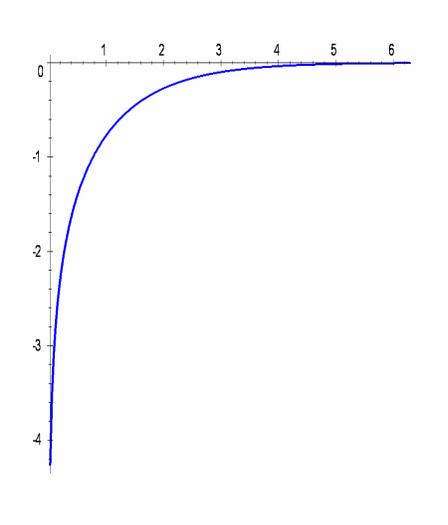,width=6cm,height=7cm}}
\caption{Fig 5.3: $\Delta E $ for the open Friedmann model which
diverges at $\protect\eta =0$} \label{}
\end{figure}

$\smallskip $\newpage

\begin{figure}
\centerline{\epsfig{file=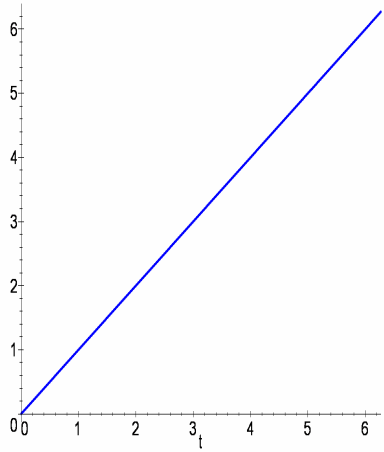,width=6cm,height=7cm}}
\caption{Fig 5.4: $\Delta E$ for the Lemaitre form of the De
Sitter universe. Here $\Lambda $ is taken to be unity.} \label{}
\end{figure}

\newpage

{\bf APPENDIX 1}

Consider the metric [1] (with signature $(+,-,-,-)$)

\begin{equation}
ds^{2}=L^{2}(u)(dx^{2}-dy^{2})-dudv  \label{A1.1}
\end{equation}

where $u=t-z\,$\thinspace and$\,v=t+z\,\,\,$which is always flat.

The metric tensor is.

\begin{equation}
g_{ab}=\left(
\begin{array}{cccc}
1 & 0 & 0 & 0 \\
0 & -L^{2}(u) & 0 & 0 \\
0 & 0 & -L^{2}(u) & 0 \\
0 & 0 & 0 & -1
\end{array}
\right)  \label{A1.2}
\end{equation}
Its inverse is.

\begin{equation}
g^{ab}=\left(
\begin{array}{cccc}
1 & 0 & 0 & 0 \\
0 & -L^{-2}(u) & 0 & 0 \\
0 & 0 & -L^{-2}(u) & 0 \\
0 & 0 & 0 & -1
\end{array}
\right)  \label{A1.3}
\end{equation}
The non vanishing Christoffel symbols are:

\begin{equation}
\left\{
\begin{array}{c}
1 \\
1\,\,\,\,0
\end{array}
\right\} =\left\{
\begin{array}{c}
2 \\
2\,\,\,\,0
\end{array}
\right\} =\left\{
\begin{array}{c}
1 \\
0\,\,\,\,1
\end{array}
\right\} =\frac{L^{\prime }(u)}{L(u)\,}.\,\,\,\,\,\,\,\,\,\,\,\,\,\,\,\,\,\,%
\,\,\,\,\,\,\,\,\,\,\,\,  \label{A1.4}
\end{equation}

\begin{equation}
\left\{
\begin{array}{c}
0 \\
1\,\,\,\,1
\end{array}
\right\} =\left\{
\begin{array}{c}
3 \\
1\,\,\,\,1
\end{array}
\right\} =L(u)L^{\prime
}(u).\,\,\,\,\,\,\,\,\,\,\,\,\,\,\,\,\,\,\,\,\,\,\,\,\,\,\,\,\,\,\,\,\,\,\,%
\,\,\,\,\,\,\,\,\,\,\,\,\,\,\,\,\,\,\,\,\,\,\,\,  \label{A1.5}
\end{equation}

\begin{equation}
\,\,\left\{
\begin{array}{c}
1 \\
3\,\,\,\,1
\end{array}
\right\} =\left\{
\begin{array}{c}
2 \\
0\,\,\,\,2
\end{array}
\right\} =\frac{L^{\prime }(u)}{L(u)\,}\,.\,\,\,\,\,\,\,\,\,\,\,\,\,\,\,\,\,%
\,\,\,\,\,\,\,\,\,\,\,\,\,\,\,\,\,\,\,\,\,\,\,\,\,\,\,\,\,\,\,\,\,\,\,\,\,\,%
\,\,\,\,\,\,\,\,\,\,\,\,\,\,\,  \label{A1.6}
\end{equation}

\begin{equation}
\,\left\{
\begin{array}{c}
1 \\
2\,\,\,\,2
\end{array}
\right\} =\left\{
\begin{array}{c}
3 \\
2\,\,\,\,2
\end{array}
\right\} =L(u)L^{\prime
}(u)\,.\,\,\,\,\,\,\,\,\,\,\,\,\,\,\,\,\,\,\,\,\,\,\,\,\,\,\,\,\,\,\,\,\,\,%
\,\,\,\,\,\,\,\,\,\,\,\,\,\,\,\,\,\,\,\,\,\,\,\,\,\,  \label{A1.7}
\end{equation}

\begin{equation}
\,\left\{
\begin{array}{c}
2 \\
3\,\,\,\,2
\end{array}
\right\} =\left\{
\begin{array}{c}
1 \\
1\,\,\,\,3
\end{array}
\right\} =\left\{
\begin{array}{c}
2 \\
2\,\,\,\,3
\end{array}
\right\} =\frac{L^{\prime }(u)}{L(u)\,}\,.\,\,\,\,\,\,\,\,\,\,\,\,\,\,\,\,\,%
\,\,\,\,\,\,\,\,\,\,\,\,\,\,  \label{A1.8}
\end{equation}
The primes refers to differentiation with respect to $u$. The nonvanishing
components of the Ricci tensor are $R_{00},\,R_{03},\,R_{30}\,$and$%
\,\,R_{33}.$ Using Christoffel symbols listed above we obtain the following
Ricci tensor components.

\begin{equation}
R_{00}=R_{33}=\frac{-2L^{\prime \prime }(u)}{L(u)}\,.\,\,\,\,\,\,\,\,\,\,\,%
\,\,\,\,\,\,\,\,\,\,\,\,\,\,\,\,\,\,\,\,\,\,\,\,\,\,\,\,\,\,\,\,\,\,\,\,\,\,%
\,\,\,\,\,\,\,\,\,\,\,\,\,\,\,\,\,\,\,\,\,\,\,\,\,\,\,\,\,\,\,
\label{A1.9}
\end{equation}

\begin{equation}
R_{30}=R_{03}=\frac{2L^{\prime \prime }(u)}{L(u)}\,.\,\,\,\,\,\,\,\,\,\,\,\,%
\,\,\,\,\,\,\,\,\,\,\,\,\,\,\,\,\,\,\,\,\,\,\,\,\,\,\,\,\,\,\,\,\,\,\,\,\,\,%
\,\,\,\,\,\,\,\,\,\,\,\,\,\,\,\,\,\,\,\,\,\,\,\,\,\,\,\,\,\,\,\,\,
\label{A1.10}
\end{equation}
From equations $\left( A1.9\right) $ and $\left( A1.10\right) $ the required
result follows.

\newpage

\smallskip

\smallskip

{\bf APPENDIX 2}

We shall prove here that
\begin{equation}
F^{0}=m[(\ln A)_{,0}-\Gamma _{00}^{0}+\Gamma _{0j}^{i}\Gamma
_{0i}^{j}/A]f^{2},\,\,\,\,\,\,\,\,\,\,\,\,\,\,\,\,  \label{A2.1}
\end{equation}
and

\begin{equation}
F^{i}=\Gamma
_{00}^{i}f^{2}\,\,\,\,\,\,\,\,\,\,\,\,\,\,\,\,\,\,\,\,\,\,\,\,\,\,\,\,\,\,\,%
\,\,\,\,\,\,\,\,\,\,\,\,\,\,\,\,\,\,\,\,\,\,\,\,\,\,\,\,\,\,\,\,\,\,\,\,\,\,%
\,\,\,\,\,\,\,\,\,\,\,\,\,\,\,\,\,\,\,\,\,\,\,\,  \label{A2.2}
\end{equation}
are the solutions of the following equations.
\begin{equation}
l^{i}(F_{,i}^{0}+\Gamma
_{ij}^{0}F^{j})=0,\,\,\,\,\,\,\,\,\,\,\,\,\,\,\,\,\,\,\,\,\,\,\,\,\,\,\,\,\,%
\,\,\,\,\,\,\,\,\,\,\,\,\,\,\,\,\,\,\,\,\,\,\,\,\,\,\,\,\,\,\,\,\,\,\,\,\,\,%
\,\,\,\,\,  \label{A2.3}
\end{equation}

\begin{equation}
l^{j}(F_{,j}^{i}+\Gamma
_{0j}^{i}F^{0})=F^{*i}.\,\,\,\,\,\,\,\,\,\,\,\,\,\,\,\,\,\,\,\,\,\,\,\,\,\,%
\,\,\,\,\,\,\,\,\,\,\,\,\,\,\,\,\,\,\,\,\,\,\,\,\,\,\,\,\,\,\,\,\,\,\,\,\,\,%
\,\,\,\,\,\,\,  \label{A2.4}
\end{equation}
where
\begin{equation}
F^{*i}=m(\Gamma _{00,j}^{i}-\Gamma _{0j,0}^{i}+\Gamma
_{0j}^{i}\Gamma _{00}^{0}-\Gamma _{0k}^{i}\Gamma
_{0}^{k})f^{2}l^{j}.  \label{A2.5}
\end{equation}
For the verification of equation $\left( A2.3\right) $ we note that
\begin{equation}
A=\Gamma
_{0i}^{i},\,\,\,\,\,\,\,\,\,\,\,\,\,\,\,\,\,\,\,\,\,\,\,\,\,\,\,\,\,\,f^{2}=%
\frac{1}{g_{00}}.  \label{A2.6}
\end{equation}
Thus
\begin{equation}
(\ln A)_{,0}=g_{,0}^{ij}/g^{ij}+g_{ij,00}/g_{ij,0}.  \label{A2.7}
\end{equation}
Here we will make use of the Riemann normal coordinates (RNCs) for spatial
directions. We have
\begin{equation}
g_{0i}=0  \label{A2.8}
\end{equation}
\begin{equation}
g_{\mu \nu ,0}\neq 0,\,\,\,g_{\mu \nu ,i}=0=g_{\mu \nu ,0i}.
\label{A2.9}
\end{equation}
Using this approximation we can write
\begin{equation}
(\ln A)_{,0i}=0=\Gamma _{00,i}^{0}=\Gamma _{0j}^{i}\Gamma
_{0i,i}^{j}. \label{A2.10}
\end{equation}
Now
\begin{eqnarray}
F_{,i}^{0} &=&m[(\ln A)_{,0}i-\Gamma _{00,i}^{0}+(\Gamma _{0j}^{i}\Gamma
_{0i}^{j}/A)_{,i}]f^{2}  \nonumber \\
&&+m[(\ln A)_{,0}-\Gamma _{00}^{0}+\Gamma _{0j}^{i}\Gamma
_{0i}^{j}/A](f^{2})_{,i}.  \label{A2.11}
\end{eqnarray}
Using RNCs the first, second, third and the last term on the right hand side
vanishes. Therefore
\begin{equation}
F_{,i}^{0}=0  \label{A2.12}
\end{equation}
Thus equation $\left( A2.3\right) $ is satisfied. Equation $\left(
A2.4\right) $ can directly be obtained just by replacing the values of $%
F^{0} $ and $F^{i}$ from equations $\left( A2.1\right) $ and $\left(
A2.2\right) $.

Hence $F^{0}$ and $F^{i}$ given by equations $\left( A2.1\right) $ and $%
\left( A2.2\right) $ are the solutions of equations $\left( A2.3\right) $
and $\left( A2.4\right) $.

\smallskip \newpage

\textbf{References}
\begin{enumerate}
\item  C. W. Misner, K. S. Thorne and J. A. Wheeler, {\it Gravitation,} W.
H. Freeman,\thinspace \thinspace San Francisco, (1973).

\item  D. Kramer,\thinspace H. Stephanie, E. Herlt and McCallum, {\it Exact
solutions of Einstein's Field Equations, }(Cambridge university, Press,
Cambridge, 1979).

\item  J. Weber, {\it General Relativity and Gravitational Waves, }%
(Interscience, NewYork, 1961).

\item  \ J. Ehlers and W. Kundt,\thinspace \thinspace {\it Gravitation: An
Introduction to Current Research,} ed. L. Witten (Wiley, New York, 1962).

\item  \ J. Weber and J. A. Wheeler, Rev. Mod. Phy. {\bf 29} (1957) 509.

\item  S. M. Mahajan, A.\ Qadir, P. M. Valanju, Nuovo Cimento {\bf B65}
(1981) 404; \thinspace J. Quamar, Ph.D thesis, Quaid-i-Azam University
(1984); A. Qadir and J. Quamar,{\it \ Proceedings of the Third Marcel
Grossman Meeting on General Relativity, }ed. Hu Ning{\it \ (}Science Press
and North Holland Publishing Co. 1983) 189.

\item  \ M. Sharif, Ph.D thesis, Quaid-i-Azam University (1991).

\item  \thinspace \thinspace A. Qadir and M Sharif, Physics
Letters\thinspace A{\bf , 167} (1992) 331.

\item  \thinspace \thinspace A. Qadir and M Sharif, Nuovo Cimento
B, {\bf 107 } (1992) 1071.

\item  A. Qadir, Nuovo\thinspace \thinspace Cimento {\bf 112B} (1997) 485.

\item  M Sharif, Astrophysics and Space Science {\bf 253} (1997)

\item  Martin Rees, R Ruffini, J. A. Wheeler {\it Black Holes Gravitational
Waves and Cosmology, }(Gordon and Breech, Science Publishers Inc.)1976.

\item  A. Qadir. {\it Einstein's General Theory of Relativity,} (preprint).

\item  M. Sharif, to appear in Astrophysics and Space Science.

\item  R. Penrose and W. Rindlers,{\it \ Spinors and Spacetime, }Vol 1,
Cambridge University Press 1984.
\end{enumerate}
\end{document}